\documentclass[showpacs,amsmath,amssymb,aps,pra,
longbibliography,
10pt,reprint,superscriptaddress]{revtex4-2}
\usepackage{bm}
\usepackage[breaklinks=true,colorlinks=true,linkcolor=blue,urlcolor=blue,citecolor=blue]{hyperref}
\usepackage{dcolumn}
\usepackage{amsmath,amssymb}
\usepackage{mathdots}
\usepackage{natbib}
\usepackage{soul,color}
\usepackage{graphicx}
\usepackage{epstopdf}
\usepackage[note-name]{notes2bib}
\usepackage{mathptmx}
\usepackage{orcidlink}
\usepackage{sidecap}

\begin{document}

\title{Spin waves in magnetic nanodisks, nanorings, and 3D nanovolcanoes}

\author{Oleksandr Dobrovolskiy\,\orcidlink{0000-0002-7895-8265}}
\email[Corresponding author:]{oleksandr.dobrovolskiy@tu-braunschweig.de}
\affiliation{   Cryogenic Quantum Electronics, 
                Institute for Electrical Measurement Science and Fundamental Electrical Engineering,
                Laboratory for Emerging Nanometrology (LENA),
                Technische Universit\"at Braunschweig, 
                Braunschweig, Germany}

\author{Gleb Kakazei\,\orcidlink{0000-0001-7081-581X}}
\affiliation{   Departamento de Fisica e Astronomia, Faculdade de Ci\^{e}ncias,
                Institute of Physics for Advanced Materials, Nanotechnology and Photonics,
                Universidade do Porto, Porto, Portugal
                }

\begin{abstract}
Patterned magnetic nanostructures are advanced materials characterized by their unique magnetic properties at the nanoscale, which are the result of tailored geometric configurations and compositional engineering. As interest in nanotechnology continues to grow exponentially, the exploration of patterned magnetic nanostructures turns into a vibrant and critical area of study for both industry professionals and academic researchers. Here, we review investigations of standing spin waves (collective spin precessions) in magnetic elements by the technique of ferromagnetic resonance (FMR). The presentation encompasses earlier studies of arrays of magnetic nanodisks and nanoringes by cavity-based FMR as well as more recent studies of individual nanodisks and 3D nanovolcanoes by a broadband FMR method which implies the use of a coplanar waveguide. Overall, the manuscript outlines the development of FMR studies along the two major lines: (i) downscaling from multiple to individual magnetic nanoelements and (ii)~extending planar nanomagnets into the third dimension.
\end{abstract}
\maketitle

\tableofcontents

\section{Introduction}

Over the past decades, the rapid advancement in magnetic recording storage capacity has underscored the limitations of existing technologies. This has sparked considerable interest in patterned nanostructures as promising candidates for next-generation magnetic storage media, placing their fabrication and characterization at the forefront of magnetism research\,\cite{Ros01rmr}. Beyond industrial applications, patterned magnetic structures serve as excellent model systems to study the fundamental physical properties of small magnetic particles. For instance, it was demonstrated that magnetostatic interactions are crucial in the magnetization reversal processes of ferromagnetic submicron dot arrays with small interdot spacing\,\cite{Gus99apl}. One key parameter in such studies is the induced in-plane anisotropy field, typically estimated from characteristic fields (nucleation and annihilation) observed in hysteresis loops along easy and hard magnetization directions.

As spintronics and high-density recording technologies advance, the demand for faster magnetic switching processes has grown. This has intensified efforts to understand spin dynamics and magnetic relaxation on nanosecond timescales. Numerous experimental investigations into spin excitations in magnetic dot arrays have been conducted (see\,\cite{Dem01inb,Bay06inb} for references). Many of these studies employ Brillouin Light Scattering\index{Brillouin light scattering} (BLS), a technique renowned for its high sensitivity and ability to yield high signal-to-noise ratios without requiring large patterned areas\,\cite{Seb15fip}. With advancements in microfocused BLS\,\cite{Mad12inb} and time-resolved scanning magneto-optical Kerr microscopy\,\cite{Per05prl}, the spin-wave eigenmode spectrum within individual micron-scale elements can now be visualized. Recently, BLS has been used to demonstrate qualitative differences in spatially resolved spin-wave resonances of 2D and 3D nanostructures, which originate from the geometrically induced non-uniformity of the internal magnetic field\,\cite{Lam23nan}.

Ferromagnetic resonance\index{ferromagnetic resonance} (FMR) has established itself as a highly effective technique for investigating the magnetic properties of continuous thin films and multilayers. It enables the precise determination of exchange interactions and various types of magnetic anisotropy fields, as extensively documented in previous studies\,\cite{Hei91jap,Blo92jap,Far98rpp,Gol97mmm}. One significant advantage of high-frequency FMR (operating at 10\,GHz or above) is that the resonance field often surpasses the sample's saturation field, effectively eliminating the influence of domain structures. Additionally, the narrow linewidths achievable with FMR allow for highly accurate resonance position measurements, with uncertainties as small as a few Oersteds. This precision facilitates detailed analysis of angular dependence (both polar and azimuthal) and temperature dependence of resonance fields. FMR has also proven invaluable in exploring standing spin waves in continuous thin magnetic films, including both single-layered systems\,\cite{Wig64phr} and multilayers\,\cite{Kor93jap}.

This manuscript presents on the application of FMR to probe standing spin waves in magnetic nanodisks, nanorings and 3D nanovolcanos. It highlights the development of FMR along the two major lines of (i) downscaling the system under study from multiple to individual magnetic nanoelements and (ii) extending planar nanomagnets into the third dimension.

\section{Spin waves in arrays of nanomagnets}
\label{cArrays}

\subsection{Spin waves in arrays of nanodisks}

\begin{SCfigure*}
    \centering
    \includegraphics[width=1.36\linewidth]{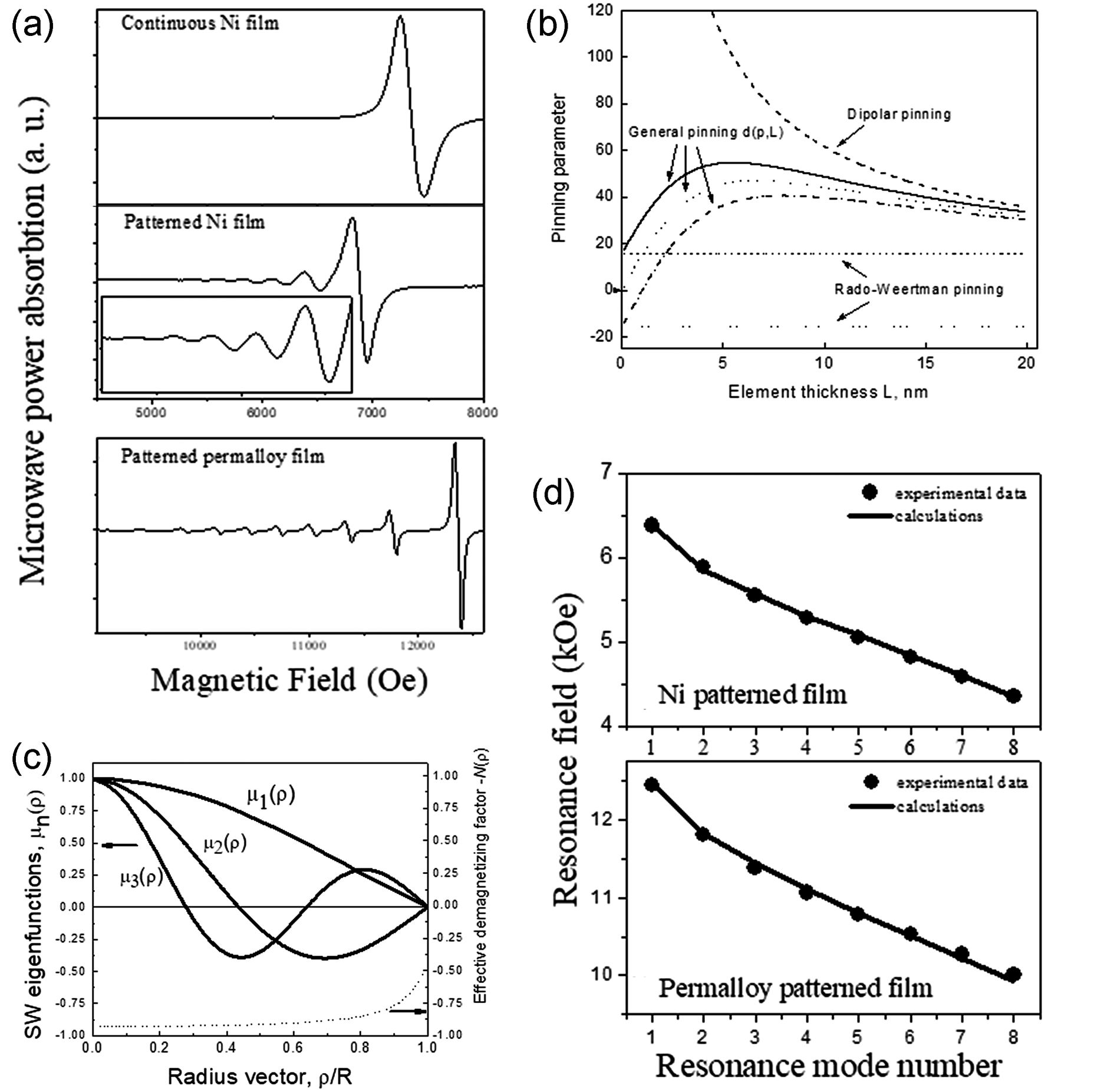}
    \caption{
    (a) FMR resonance spectra for a continuous Ni film and the arrays of Ni and Py circular dots.
    (b) General pinning parameter $d(p,L)$ for a thin magnetic disk vs disk thickness $L$ for a disk radius of $R=500$\,nm. The dashed line corresponds to pure dipolar pinning ($K_s =0$, $L_e=0$) of Ref.\,\cite{Gus02prb}. Here, $K_s$  is the constant of surface anisotropy and $L_e = \sqrt{2A/M_s^2}$ is the characteristic exchange length.
    The solid line corresponds to the easy-plane surface anisotropy ($K_s/\pi M_s^2 = -1$\,nm) while the dot-dashed line corresponds to the easy-axis type of surface anisotropy ($K_s/\pi M_s^2 = 1$\,nm), Eq.\,\eqref{e6}. The dotted line corresponds to $K_s=0$, the horizontal lines correspond to the Rado-Weertman pinning with $K_s/\pi M_s^2 = -1$\,nm (upper) and $K_s/\pi M_s^2 = 1$\,nm (lower), $L_e = 20$\,nm.
    (c) Spin-wave eigenmodes profiles for the first three modes and the coordinate-dependent effective demagnetizing factor $–N_s(\rho)$ for a cylindrical dot along the radial direction.
    (d) Comparison of measured and calculated FMR resonance peak positions: Top panel -- patterned Ni film, parameters used in the calculation are $M_s = 484$\,emu/cm$^3$, $L = 70$\,nm, $R = 500$\,nm, $H_\perp = 1.84$\,kOe, $A = 1.84 \times 10^{-7}$\, erg/cm; Bottom panel -- patterned permalloy film, parameters used in the calculation are $M_s = 830$\,emu/cm$^3$, $L = 50$\,nm, $R = 500$\,nm, $H_\perp = 0$, $A = 1.4 \times 10^{-6}$\,erg/cm. Panels (a), (c) and (d) are adapted with permission from\,\cite{Kak04apl}.}
    \label{fGleb1}
\end{SCfigure*}

An analytical description of spin waves in patterned magnetic structures is rarely feasible. Due to the inhomogeneity of the internal demagnetizing field, it is impossible to find exact eigenfunctions for most of the geometries. The good agreement between analytical calculations and experimental data was found only for few cases of simplest symmetry. One of the most known examples is long flat infinite stripes magnetized along short direction\,\cite{Jor02prl}. For this system it was natural to assume that the eigenfunctions of the stripes would have an almost sinusoidal form, analogous to the form of spin wave resonance modes in a perpendicularly magnetized continuous homogeneous film. On the contrary, the theoretical description of magnetization dynamics in tangentially magnetized circular dot arrays in terms of excited spin wave modes encounters substantial difficulties due to the absence of axial symmetry in the dot plane\,\cite{Gus00jap,Gub03jap}. Thus, it is very interesting to study experimentally the dynamic properties of \emph{perpendicularly} magnetized circular dot arrays where cylindrical symmetry is preserved, and theoretical interpretation of the observed spin wave modes is relatively simple.

Previously, the BLS technique was applied for experimental studies of high-frequency spin-wave modes in arrays of axially magnetized long magnetic nanowires \cite{Wan02prl} having a large aspect ratio (height/radius) $p = L/R \simeq 50$. Three spin wave modes, caused by the finite radius of the wire, were observed and fitted using the modified theory by Arias and Mills\,\cite{Ari01prb}.

In this section, we outline the results of studies of spin-wave\index{spin waves} modes by the FMR technique in the X-band ($7–11.2$\,GHz) in the opposite limiting case -- in perpendicularly magnetized thin circular magnetic dots with $p = L/R\simeq0.1$ and a radius $R = 500$\,nm. Circular Ni dots with $L = 50$\,nm were arranged into rectangular arrays with a fixed interdot distance of $1000$\,nm along one lattice axis and variable distances from $50$ to $800$ nm along the other axis. At a later stage, an additional array of non-interacting circular permalloy (Ni$_{81}$Fe$_{19}$) dots with $L=50$\,nm and interdot distance $2.5\,\mu$m was fabricated to confirm the theoretical model. The total patterned area of each sample was $4$\,mm$\times4$\,mm. Room temperature FMR spectra were recorded at the frequency of $9.3$\,GHz (Ni) and $9.85$\,GHz (Ni$_{81}$Fe$_{19}$) using cavity-based electron spin resonance spectrometer. The external magnetic field was applied along the normal of the film ($\mathbf{H}\parallel \mathbf{n}$). The saturation magnetization of the dot arrays and the continuous films was measured using a SQUID magnetometer.

When the applied field was close to the sample normal, multiple sharp resonant peaks (up to $8$ for permalloy dots) were observed in the field region below the main FMR peak [central and bottom panels of Fig.\,\ref{fGleb1}(a)]. No sight of such periodic spectra was found for the reference Ni [top panel of Fig.\,\ref{fGleb1}(a)] and permalloy continuous films, supporting the idea that the additional modes with discrete frequencies are due to the confined in-plane geometry of the patterned films and are of magnetostatic origin.

For all samples of the same material the relative positions of individual resonance peaks are independent of the separation between the dots in the array while the absolute field positions of the peaks change with the change of the dot separation. This property of the FMR spectra in perpendicularly magnetized dot arrays indicates that the dipole–dipole interaction between the dots creates an additional effective perpendicular bias field, but it does not change the structure of the spectrum of spin wave modes.

To explain the experimentally observed multi-resonance FMR spectra, a dipole-exchange theory was developed for spin-wave spectra in an infinite in-plane magnetic film where formation of standing spin-wave modes is caused by the quantization of the in-plane component of the wave vector due to the finite dot radius. For a perpendicularly magnetized infinite in-plane magnetic film of the thickness $L$, an approximate diagonal dipole-exchange spin-wave dispersion equation (see Ref.\,\cite{Kal86jpc,Kal90pcm}) can be written in a form similar to the classical Herring–Kittel spin-wave dispersion equation in a bulk sample\,\cite{Her51phr}. For the lowest spin-wave mode, that is assumed to be uniform along the film thickness, the dispersion equation reads
\begin{equation}
    \label{e1}
    \omega_k^2= [\omega_H +\alpha \omega_M k^2][\omega_H+\alpha \omega_M k^2+\omega_M f(kL)],
\end{equation}
where $\omega_H= \gamma H_i$, $\omega_M =  \gamma 4\pi M_s$, $H_i = H_e - 4\pi M_s + H_\perp$ is the effective internal bias magnetic field with the external field $H_e$ and the saturation magnetization $M_s$, $H_\perp$ the perpendicular anisotropy field, $\gamma$ the gyromagnetic ratio, $\alpha = A/2\pi M_s^2$ the exchange constant expressed in cm$^2$, $A$ the exchange stiffness constant in erg/cm, and $k$ the modulus of the in-plane spin-wave wave vector. The matrix element $f(kL)$ of the dipole–dipole interaction for a perpendicularly magnetized film which replaces $\sin^2\theta_k$ in the classical Herring–Kittel dispersion equation for an infinite ferromagnetic medium has the following form\,\cite{Kal86jpc,Kal90pcm}:
\begin{equation}
    \label{e2}
    f(kL)=1-\frac{1-\exp(-kL)}{kL}.
\end{equation}
Equation\,\eqref{e1} was derived for plane spin waves. However, it can be used to approximately describe the frequencies of standing in-plane spin-wave modes in a magnetic dot having the shape of a thin disk $L/R \ll 1$. The finite in-plane size $R$ of the dot brings two qualitative features into the dispersion equation. First, due to the non-ellipsoidal shape of the dot the demagnetizing field inside the dot becomes inhomogeneous and the internal bias field depends on the radial coordinate $\rho$ (see Ref.\,\cite{Jos65jap} for details),
\begin{equation}
    \label{e3}
    H_i(\rho)=H_e-4\pi M_sN(\rho)+H_\bot,
\end{equation}
where the coordinate-dependent effective demagnetizing factor $N(\rho)$ for a cylindrical dot along the normal direction is defined by Eq.\,(38) in Ref.\,\cite{Jos65jap}.

Second, due to the finite radius of the disk-shaped dot, only discrete values of the in-plane wave vector are allowed $k\rightarrow k_m$, where $m = 1, 2, 3,\dots$ is the radial mode number, i.\,e., standing spin-wave modes in dots have quantized values of the in-plane wave number.

In cylindrical dots for the description of spatial profiles of dipole-exchange standing spin wave modes it is possible to use the radial eigenfunctions of the exchange differential operator of the second order that have the form\,\cite{Gus00mmm}
\begin{equation}
    \label{e4}
    \mu_n(\rho)=M_sJ_n(\kappa_{nm}\rho).
\end{equation}
where $J_n(x)$ is the Bessel function with integer index $n=0, \pm1, \pm2 \dots$ and $\kappa_{n,m} = \alpha_{n,m}/R$ is the quantized radial wave number with $\alpha_{n,m}$ being the $n$-th root of the equation $J_n^\prime(x) + d(p,L)J_n(x)=0$ with the general pinning parameter $d(p,L)$. The functions given by Eq.\,\eqref{e4} should satisfy the general boundary condition at the dot lateral boundary $\rho=R$\,\cite{Gus05prb}
\begin{equation}
    \label{e5}
    R\frac{d\mu_n(k_{nm}\rho)}{d\rho}+d(p,L)\mu_n(k_{nm}\rho)\mid_{\rho=R}=0,
\end{equation}
where the effective pinning parameter at the dot lateral boundary $d$ is determined by the expression
\begin{equation}
    \label{e6}
    d(p,L)=\frac{2\pi[1-(\frac{K_s}{\pi M_s^2L})]}{p[b_1+b_2\ln(1/p)+(\frac{L_e}{L})^2]}.
\end{equation}
Here $K_s$  the constant of surface anisotropy at the lateral boundary of the dot, $L_e = \sqrt{2A/M_s^2}$ the characteristic exchange length of a material (defining the length scale at which the exchange interaction becomes important), $A$ the exchange stiffness, and the dipolar constants $b_1$ and $b_2$ in the cylindrical geometry have the values $b_1 = 2(6\ln2 - 1) \approx 6.32$ and $b_2 = 4$\,\cite{Gus05prb}.

Figure\,\ref{fGleb1}(b) illustrates these functions for a thin ($L< L_e$) cylindrical dot where the pinning at the dot lateral boundary is determined, for the most part, by the surface anisotropy. Thus, in thin dots made of a magnetically soft and isotropic material (like permalloy) the edge pinning of variable magnetization could be very low and the spin-wave modes can satisfy the condition $d\mu(k_{n,m}/\rho)/d\rho = 0$  of ``free'' spins at the lateral boundary $\rho = R$.

On the other hand, for sufficiently thick ($L> L_e$) cylindrical magnetic dots with enough small aspect ratios $p = L/R\ll 1$, the effective pinning at the dot lateral boundary is practically independent of the constant of surface anisotropy, has a dipolar nature (see\,\cite{Gus05prb} and\,\cite{Gus02prb} for details), and is rather strong. In such a case it is natural to assume that the dipole-exchange eigenmodes of a thin disk-like dot excited by FMR having the shape of a zeroth-order Bessel function
\begin{equation}
    \label{e7}
    \mu_0(\rho)=M_sJ_0(k_{0m}\rho)
\end{equation}
will satisfy the boundary conditions of ``total pinning'' $\mu(\rho = R) = 0$ at the dot lateral boundary. In such a case the quantized values of the in-plane wave vector $k_m$ will be determined by the equation $k_m = \beta_{0m}/R$, where $\beta_{0m}$ are the roots of the zeroth-order Bessel function $J_0(\beta_{0m}) = 0$.

The ``pinned'' standing mode profiles by Eq.\,\eqref{e7} for the first three spin-wave modes $m=1, 2, 3$ in the dot are presented in Fig.\,\ref{fGleb1}(c). In the same plot, the coordinate-dependent effective demagnetizing factor $N(\rho)$ for a cylindrical dot along the normal direction is shown by a dotted line. Since different standing modes in the dot have different radial profiles (Eq.\,\eqref{e3}) and the demagnetizing field inside the dot (Eq. \,\eqref{e2}) is inhomogeneous (see Eq.\,\eqref{e4}), the effective internal bias magnetic field will be different for different modes $H_i \rightarrow H_{im}$. This mode-dependent effective internal field can be evaluated as in Ref.\,\cite{Kak03jap}
\begin{equation}
    \label{e8}
    H_{im}=H_e-4\pi M_sN_m+H_\bot
\end{equation}
using Eqs.\,\eqref{e3} and\,\eqref{e7} to calculate the effective matrix elements of the inhomogeneous demagnetizing field $N_{m}$ for different standing spin wave modes
\begin{equation}
    \label{e9}
    N_m=\frac{1}{A_m}\int_0^R N(\rho)J_0^2(k_{0m}\rho)\rho d\rho,
    \qquad
    A_m=\frac{1}{2}R^2J_1^2(k_{0m}).
\end{equation}
Substituting $k=k_{0m}= \beta_{0m}/R$ for the quantized value of the in-plane wave vector and Eq.\,\eqref{e8} for the discrete values of $H_i=H_{im}$ to the dispersion Eq.\,\eqref{e1}, and taking into account experimental microwave frequencies $9.3$ GHz for nickel and $9.85$ GHz for permalloy, resonance values of the external bias magnetic field $H_e^{res}(m)$ corresponding to different resonant spin wave modes described by Eq.\,\eqref{e7} were calculated using the following parameters: for nickel dots $M_s = 484$\,emu/cm$^3$, $L = 70$\,nm, $R = 500$\,nm, $H_\perp = 1.84$\,kOe, $A = 1.84 \times 10^{-7}$\,erg/cm; for permalloy dots $M_s = 830$\,emu/cm$^3$, $L = 50$\,nm, $R = 500$\,nm, $H_\perp = 0$, $A = 1.4 \times 10^{-6}$\,erg/cm. Note that Ni exhibits noticeable perpendicular anisotropy of magnetoelastic nature (see Ref.\,\cite{Kak96mmm}). Also the standard values of the gyromagnetic ratio $\gamma/2\pi=3.05$\,MHz/Oe for nickel and $2.96$\,MHz/Oe for permalloy\index{magnonic materials} were used. The results of comparison of the calculated and measured resonance fields for nickel and permalloy dots are presented in Fig.\,\ref{fGleb1}(d). It is clear from the figure that the described theory gives an excellent quantitative description of the experiment for both materials for all the modes including higher-order ones.

Summarizing, the multiresonance FMR spectra of laterally confined spin-wave modes in nickel and permalloy circular dot arrays are quantitatively described by a dipole-exchange theory of spin-wave dispersion in perpendicularly magnetized films. This theory accounts for the quantization of the in-plane wave vector due to the finite radius of the dots and includs an averaging of the inhomogeneous internal bias field for various spatial profiles of standing spin-wave modes. Although dipole–dipole interactions between dots, significant at small interdot separations, shifts the spectra as a whole, they do not alter the relative positions of the resonance peaks or the structure of the spectra.

\subsection{Spin waves under axial symmetry violation}
\label{cSplit}

\begin{SCfigure*}
    \centering
    \includegraphics[width=1.5\linewidth]{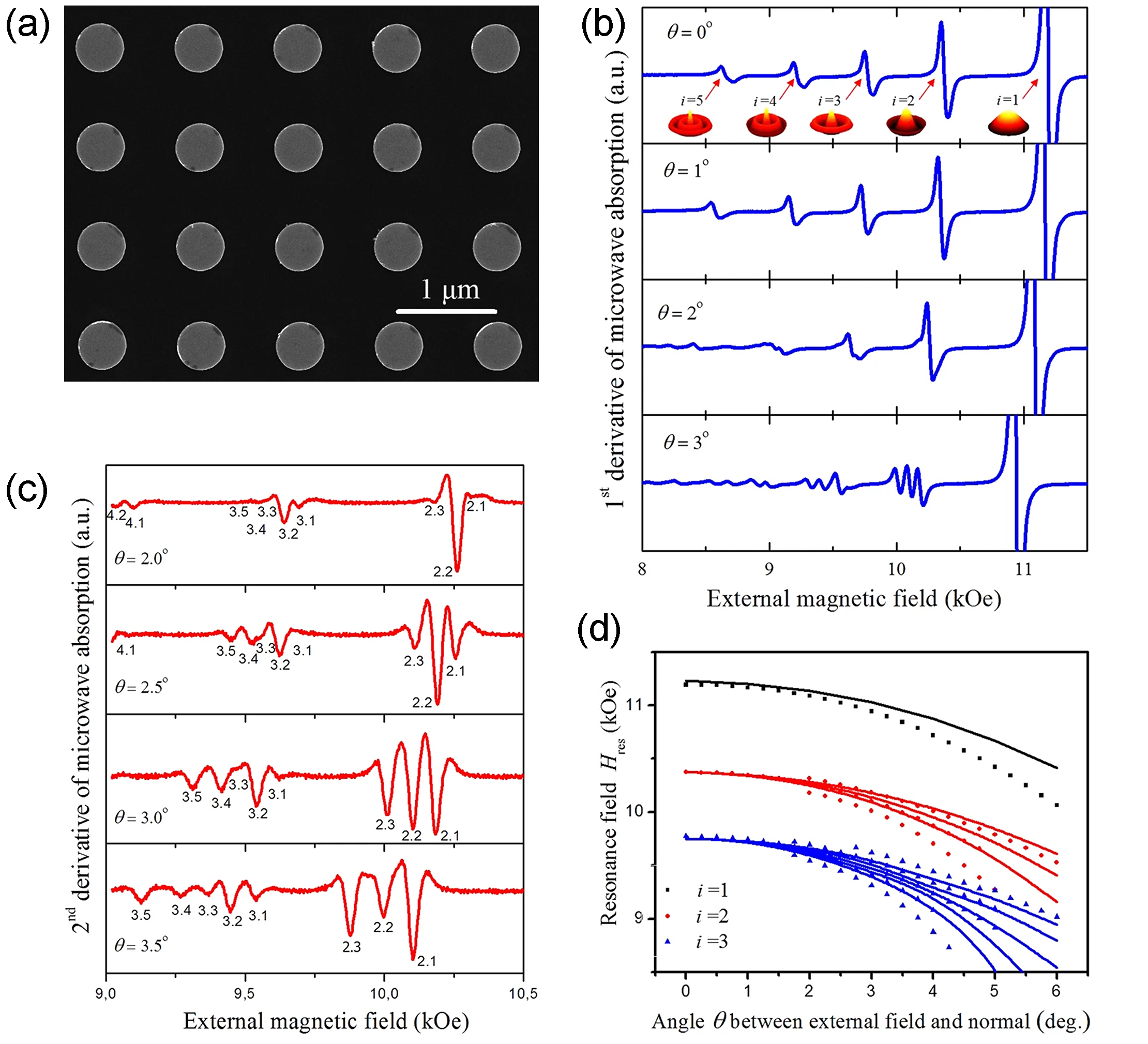}
    \caption{
    a) SEM image of the permalloy dot array under study. The dot radius is $250$\,nm, center-to-center interdot distance is $1000$\,nm and dot thickness is $40$\,nm.
    (b) Evolution of the 1st derivative of the spin-wave spectra of an array of circular dots with increase of angle $\theta$. The main mode with index $i = 1$ was partially cut to make all other modes clearly visible. Inserts show zero-order Bessel-function profiles for corresponding circular ``drumhead'' modes.
    (c) Evolution of the 2nd derivative of the spin-wave spectra of an array of circular dots with increasing angle between the normal to the sample surface and direction of the external magnetic field. The peaks of the spin wave modes are numbered as $i.s$, where $i$ is the mode number and $s$ is the level number. The first mode $i = 1$ is not shown in the plot.
    (d) Dependence of the resonance fields of the spin-wave modes [numbered by (i)] on the angle between the normal to the sample surface and the direction of the external magnetic field. Points represent the experimental data while the lines are theoretical fits from Ref.\,\cite{Bun15nsr}. Adapted with permission from\,\cite{Bun15nsr}.}
    \label{fGleb2}
\end{SCfigure*}

In the previous section it was convincingly demonstrated that for perpendicularly magnetized circular dots, the profiles of standing spin-wave modes can be accurately described by zero-order Bessel functions. However, for practical applications of patterned elements, i.\,e. in read-write magnetic heads, the direction of the external magnetic field may deviate from the axial symmetry. Since in such devices additional spin-wave modes can be considered as magnetic noise\,\cite{Hel10prb}, it is important to investigate the evolution of the spin-wave spectra in the case of a symmetry violation.

Here we outline the results of studies of the modification of spectra of circular permalloy dots in external magnetic fields slightly inclined from the perpendicular direction. For this orientation, the most efficient experimental techniques are either cavity-based\,\cite{Kak04apl,Kak08jnn,Ned13jap,Cas12prb,Dou10jap} or broadband \cite{Yuc07jap} FMR spectroscopy as well as FMR force microscopy\,\cite{Mew06prb,Chi12prl}. In what follows we summarize the results of studies using the cavity-based X-band FMR technique that allows on to probe millimeter-size samples and to use sample holders with both, in-plane and out-of-plane angular rotations. Room temperature measurements were performed at $9.85$\,GHz using standard electron spin resonance spectrometers. In both cases, a computer-controlled goniometer allowed varying the angle $\theta$ between the external magnetic field $H$ and the sample normal to the array plane $n$ with an accuracy of $0.1^\circ$.

Square arrays of circular permalloy dots were fabricated on a Si wafer with an oxidized surface by means of the electron-beam lithography and lift-off technique using an electron-beam writer. Since the distance between the neighbour resonance peaks of standing spin waves in a perpendicularly magnetized circular dot is inversely proportional to the dot radius\,\cite{Kak04apl}, $R = 250$\,nm was selected. In this case the shape of the dot is very close to the circular, as confirmed by scanning electron microscopy (SEM) after removing the resist, see Fig.\,\ref{fGleb2}(a), and the distance between the resonance peaks is large enough to study the splitting in a broader angle range. The interdot center-to-center distance was equal to $4R = 1000$\,nm, which is sufficient to exclude the influence of interelement magnetostatic interactions on the resonance peak positions\,\cite{Kak06prb}. The total size of the array was $2000\times2000$ elements, appropriate to obtain spin-wave spectra with a signal-to-noise ratio $>100$ even for the 5th peak. After the resist development, $40$\,nm of permalloy was deposited using molecular-beam epitaxy in a vacuum of $10^{–8}$\,mbar. To improve the magnetic material quality, a $2$\,nm thick Cr underlayer was deposited on the substrate prior to the permalloy growth. To protect the film, it was covered by a $2$\,nm thick Al layer, that fully oxidized after exposure to the atmosphere. Finally, after lift-off process the array of circular permalloy dots with well-pronounced edges was formed.

FMR measurements were started with a precise orientation of the array under study with respect to the external magnetic field. To do so, the angular dependence of the resonance field for the most intense peak was measured in the vicinity of the normal as a function of $\theta$. The maximum value of $H_{res}$ corresponded to the perfect alignment of $\mathbf{n}$ along $\mathbf{H}$. For this geometry, five sharp resonance peaks with their intensities growing with increasing resonance field were clearly observed (see Fig.\,\ref{fGleb2}(b), $\theta = 0^\circ$). This spectrum is similar to the one observed in the previous section and in Ref.\,\cite{Kak04apl}, where profiles of these standing spin waves were described well by zero-order Bessel functions. Resonance fields and interpeak distances are a bit different from those observed in\,\cite{Kak04apl} since they depend on the dot radius and thickness.

Then a series of spin-wave spectra were measured from $\theta = 0^\circ$ to $5^\circ$ with an increment of $0.25^\circ$. It can be assumed from Fig.\,\ref{fGleb2}(b) that the modes are starting to split with angle $\theta$ inclination from the normal, however, the features of the splitting process are not pronounced enough. Therefore, to clarify the number of split peaks for each particular mode and to track their resonance fields, the following procedure was introduced. First, the measured FMR spectrum (it is the 1st derivative of the microwave absorption) was integrated, then the 1st order low-pass Butterworth filter \cite{But30wwe} was applied to correct the signal baseline, and finally the 2nd derivative was calculated to increase the accuracy of the resonant field determination of the split modes. As a result, the following numbers of split peaks was obtained: three for the second mode and five for the third one. Starting from the 4th mode the separation process is affected by the overlap between peaks of the several modes. Fig.\,\ref{fGleb2}(c) shows the obtained dependences on $\theta$ of the resonance fields of the spin-wave modes (numbered as $i,s$; $i$: the mode number ($i = 1,2,\dots$), $s$: the level number).

The results shown in Fig.\,\ref{fGleb2}(b) and\,\ref{fGleb2}(c) can be summarized as follows: i) If $\theta\neq0$, all resonance peaks except for the main one are split, i.\,e. all modes are split even for the very small angles $\theta$ but it becomes evident only when the splitting distance is large enough. For small angles, the splitting looks just like broadening of the resonance line; ii) the splitting value is proportional to the mode number and angle $\theta$; iii) the number of the split peaks depends on the mode number as $2i–1$.

To explain the observed behavior, the extension of the analytical theory presented in the previous section was proposed for the case when the external magnetic field deviates from the perpendicular direction. In perfectly symmetric, perpendicularly magnetized dots, spin wave modes are well-described by Bessel functions. However, when the field tilts away from the normal, this symmetry is broken, and spin-wave modes begin to split. The degree of splitting depends on the mode number and the tilt angle of the magnetic field. To model this behavior, the Landau-Lifshitz equation was used, capturing the dynamics of magnetization components. The tilted field introduces non-diagonal interactions between spin-wave modes, leading to coupling and splitting. A perturbative approach was applied to describe this effect for small tilt angles. Analytical calculations reveal that the lowest mode ($i-1$) does not split because it lacks neighboring modes with which to couple. However, higher modes exhibit systematic splitting. The number of split levels is determined by the relationship $2i-1$, meaning the second mode splits into three levels, the third mode into five, and so on. The magnitude of the splitting increases with both the mode number and the tilt angle.

The profile of each mode is a combination of Bessel functions of even order, leading to the mode splitting and the unique behavior observed with increasing tilt. This systematic splitting pattern allows for the prediction of spin-wave behavior in non-perpendicular fields, with practical implications for spintronic applications. For instance, such modes can be used for signal processing or as tunable components in magnetic devices. Despite its utility, the theory is limited to small tilt angles ($\theta < 5^\circ$), beyond which more advanced modeling is required. Additionally, the effect of boundary conditions, such as magnetization pinning, introduces complexities that are not fully captured by this simplified approach. For more theoretical details, please see Ref.\,\cite{Bun15nsr}.

The proposed theory can quantitatively describe the experimental results if both angles $\theta$ and equilibrium magnetization angle $\theta_0$ are small. The absence of the full coincidence between theory and experiment in Fig.\,\ref{fGleb2}(d) can be associated with the effect of the dynamical magnetization pinning that was studied in details in\,\cite{Guo13prl}. Here, for simplicity, the pinned boundary conditions were used for all cases, when it has been proven in Refs.\,\cite{Guo13prl,Kak12prb} that the dipolar boundary conditions depend on the angle between the variable magnetization and the boundary of the dot. This means that the boundary conditions depend on the value of $\theta_0$ that was not taken into account in the proposed theory. Secondly, the perturbation theory used here is valid when the energy gaps between the mode levels of different numbers are noticeable. However, as it was shown in\,\cite{Bun15nsr}, the levels of splitting modes become close to each other or even cross near $\theta\approx5^\circ$, which evidently restrict the validity of the perturbation theory for larger angles. Despite the mentioned issues, the difference between the experimental and theoretical results is below $5\%$ even for $\theta = 5^\circ$.

Summarizing, when the external magnetic field is deviated from the direction perpendicular to the dots plane, the cylindrical symmetry of the isolated dot is violated. The experimental results show that in such a case the lowest mode does not split, the second mode splits into 3 levels and the third one splits into 5 levels. The phenomenological perturbation theory developed here explains such behavior and gives a simple rule that defines the number of levels that corresponds to the mode number in the symmetrical case as $2i–1$. The profile of each mode is a combination of Bessel functions of even order.

\subsection{Spin waves in arrays of nanorings}

\textbf{Case of thin (30\,nm) rings}. Ferromagnetic rings have attracted considerable attention in the last few decades due to their unique magnetic ground states and their potential use in a range of applications such as magnetic random access memory (MRAM)\,\cite{Cas03prb,Hay06prb,Zhu03itm}, biomedical sensing \cite{Mil02apl,Lla07apl} and magnetic logics\,\cite{Bow10pcs}. By varying the inner and outer radii of the ring\,\cite{Sch08prl,Luo08nan}, its composition (number, thickness and material of layers)\,\cite{Ros08jpd,Mad08jap} and by introducing structural defects\,\cite{Kla03jap,Gie07prb}, both static and dynamic magnetic properties in rings have been shown to be modified significantly. This high tunability leads to intensive investigations of micro- and nanorings\,\cite{Liu15adm}, particularly the stability of magnetic configuration\,\cite{Mam13jap}, magnetization reversal\,\cite{Zha10prb}, and the nucleation and velocity of domain walls\,\cite{Ric16prb,Ric16pra,Oya13prb}.

Investigations of dynamic behaviors of ring elements usually focus on spin-wave spectra at remanence (vortex or onion states)\,\cite{Neu06prl,Shi14prb,Pod06prl,Gub06prl} or in the tangential geometry, where the external field is applied in the ring plane\,\cite{Gub06prl}. In this section, the spin wave spectra of nanorings with the magnetic field applied out of plane are studied experimentally, numerically, and analytically.

\begin{SCfigure*}
    \centering
    \includegraphics[width=1.4\linewidth]{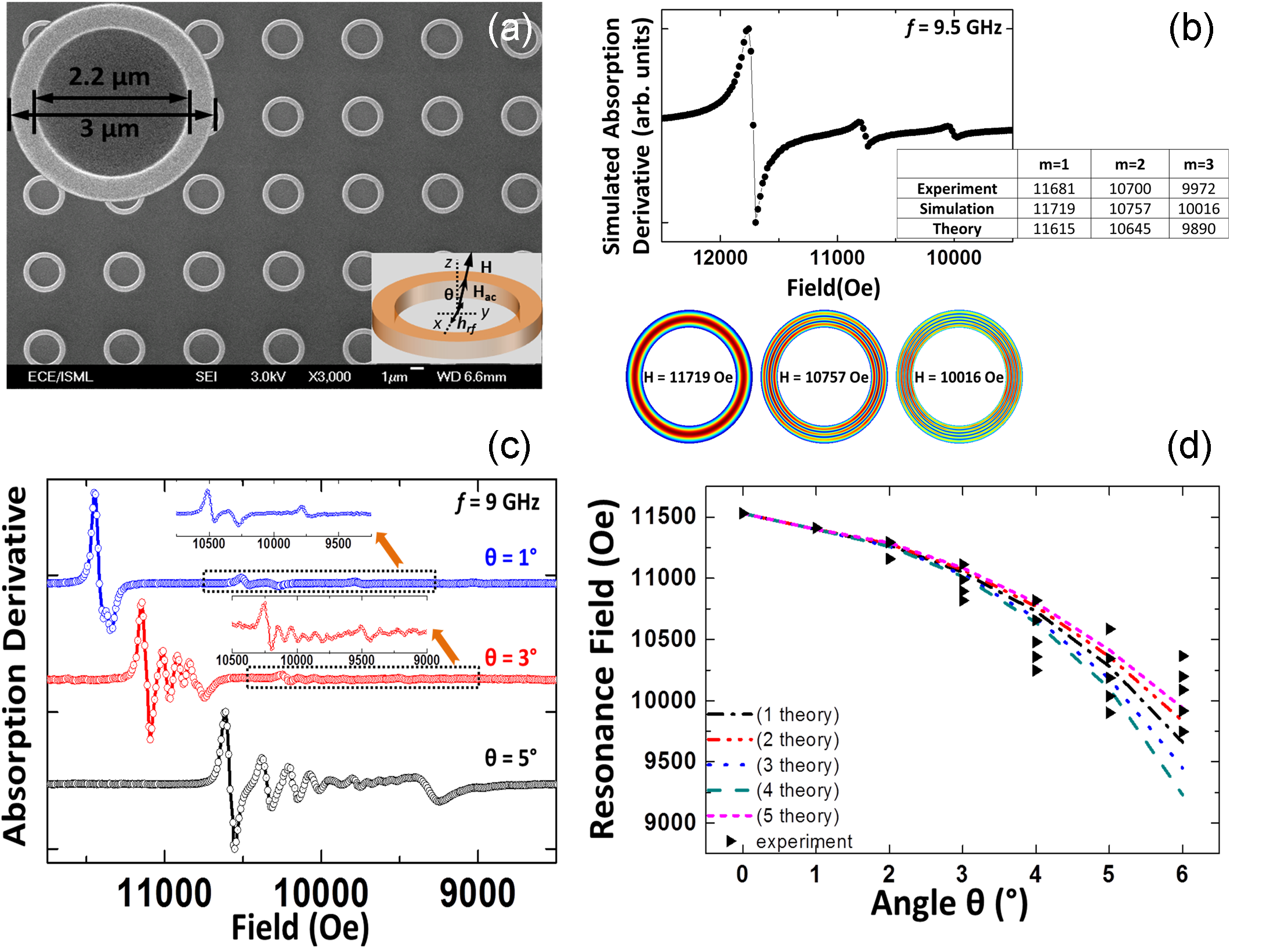}
    \caption{
    (a) SEM image of the periodic array of permalloy rings. Top left inset: SEM image of the isolated ring with indicated dimensions. Bottom right inset: the geometry of the experiment. (b) Simulation results of FMR absorption curve (upper panel: $f = 9.5$\,GHz, $\theta = 0^\circ$) and excitations profiles (lower panel) of three standing modes in the perpendicularly magnetized nanorings. The table contains the resonance fields (in Oe) at $f = 9.5$\,GHz for the three first modes: experimental data, micromagnetic simulations and theoretical results are given together for comparison.
    (c) The FMR absorption curves taken for the $30$\,nm-thick permalloy rings at $f = 9$\,GHz with $\theta$ varied from $1^\circ$ to $5^\circ$.
    (d) Resonance fields extracted from spin-wave excitations in thin ferromagnetic rings as a function of frequency for the three first modes ($m = 1, 3, 5$) in perpendicular geometry. Squares, round dots and triangles: experimental data. Solid lines: calculations by Eqs.\,(6)--(7) from Ref.\,\cite{Zho17prb}. Adapted with permission from \cite{Zho17prb}.}
    \label{fGleb3}
\end{SCfigure*}

The thin permalloy rings were fabricated using deep ultraviolet lithography, followed by electron beam evaporation and ultrasonic-assisted lift-off processes. A $30$\,nm permalloy film was deposited at a rate of $0.2$\,{\AA}/s under a high vacuum (base pressure of $2\time 10^{-8}$\,Torr). The rings were designed with an outer radius of $1500$\,nm and an inner radius of $1100$\,nm, resulting in a rim width of $400$\,nm. This geometry was selected to ensure well-defined confinement effects and to minimize fabrication irregularities. SEM images reveal good morphology and uniformity across the array. The rings exhibited sharp edges and consistent dimensions, see Fig.\,\ref{fGleb3}(a).

Broadband perpendicular FMR spectroscopy was utilized to study the spin-wave dynamics of the thin rings. The experimental setup included a $50\,\Omega$ microstrip line placed beneath the sample, through which a continuous microwave signal was applied. A $20$\,dBm microwave signal was generated by a continuous-wave microwave generator at a specific frequency between $6$ GHz and $13$\,GHz. An external magnetic bias field was carefully aligned perpendicular to the plane of the rings. The field was swept decrementally from $18$\,kOe to $0$, and additional measurements were performed by tilting the bias field by angles ranging from $0^\circ$ to $6^\circ$. A sketch of the field geometry is shown in the bottom right inset of Fig.\,\ref{fGleb3}(a). The alternating field ($\pm20$\,Oe) generated by Helmholtz coils ensured high-resolution detection of resonance peaks. The detected signal represented the first derivative of the absorption curve, providing precise insights into resonance conditions.

When the external magnetic field was applied perpendicular to the ring plane, distinct resonance peaks were observed, each corresponding to a specific spin wave mode. Micromagnetic simulations, performed using the LLG simulator, complemented the experiments. Standard permalloy parameters were employed, including a gyromagnetic ratio of $2.9$\,MHz/Oe and an exchange constant of $1.3 \time 10^{-6}$\,erg/cm. These simulations reproduced the experimental conditions, offering detailed visualizations of the mode profiles and the field dependences (Fig.\,\ref{fGleb3}(b). The resonance modes were identified as radial standing waves, with their profiles determined by the ring geometry. The fundamental mode, characterized by the simplest radial structure, exhibited the highest intensity. Higher-order modes, with more complex radial profiles, showed reduced intensities due to increased damping and spatial complexity.

The experimental results revealed a well-defined relationship between the resonance frequency and the external field. This relationship aligns with theoretical predictions\,\cite{Kak04apl}, which account for the radial quantization of spin waves in circular geometries. The spacing between the modes highlighted the influence of the ring’s inner boundary, a feature absent in simpler disk-like structures (previous sections and Ref.\,\cite{Kak04apl}).

When the external field was slightly canted from the normal, the profile of the absorption curve was significantly modified. Not only radially but azimuthally quantized spin wave modes were observed. Shown in Fig.\,\ref{fGleb3}(c) are the FMR absorption curves taken for $f = 9$\,GHz with $\theta$ varied from $1^\circ$ to $5^\circ$ with an increment of $2^\circ$. Different from the FMR curve observed at $\theta = 0^\circ$, the first two modes appearing at lower resonance fields ($H_{1-1} = 11.41$\,kOe, $H_{1-2} = 10.49$\,kOe), start to split for $\theta = 1^\circ$ and the splits become more obvious with the increase in the inclination. As the external field deviates further ($\theta = 3^\circ$), the resonance fields decrease drastically ($H_{3-1} = 11.11$\,kOe, $H_{3-2} = 10.22$\,kOe) while the number of splits increases. For $\theta = 3^\circ$ four splits for the first mode and more than five splits for the second mode were observed. As the field was tilted by $5^\circ$, the resonance mode is shifted downwards by $520$\,Oe ($H_{5-1} = 10.59$\,kOe) compared with $\theta = 3^\circ$. It was also noticed that for $\theta = 5^\circ$ the number of splits for the first mode increased to five and the field gaps in between the splitting modes get larger. Taking the field difference in between the first two splits for the first mode for an example, the field gap increases from $124$\,Oe ($\theta = 3^\circ$) to $239$\,Oe ($\theta = 5^\circ$).

Micromagnetic simulations provided further insights into this behavior, revealing that tilting the field modifies the effective demagnetizing factors within the ring, creating azimuthal variations in the internal field distribution. These variations drive the observed mode splitting, emphasizing the sensitivity of spin-wave dynamics to external perturbations.

The spin-wave modes in thin rings were analyzed using a modified dispersion relation (please see Ref.\,\cite{Zho17prb} for more details), which incorporates the effects of radial confinement and boundary conditions. Theoretical models predicted the observed quantization of radial modes and the symmetry-breaking effects of tilted fields. By coupling analytical calculations with micromagnetic simulations, the study provided a comprehensive understanding of the interplay between geometry and magnetic interactions.

The breaking of circular symmetry due to the tilted field allowed for the interaction of azimuthally distinct modes. These interactions led to a redistribution of mode intensities and shifts in resonance fields, both of which were accurately captured by the theoretical framework. Results for excitations with $f = 9$\,GHz are shown in the Fig.\,\ref{fGleb3}(d). The difference between theoretical results and experimental data can be explained by the change of boundary conditions in the canted geometry\,\cite{Kak12prb}, which was not taken into account. Still, the agreement between theory, simulation, and experiment features the robustness of the approach.

The results obtained for thin rings underscore the importance of symmetry-breaking effects in shaping spin wave dynamics. Unlike planar or disk-like structures, thin rings exhibit more complicated behaviors due to their circular geometry and inner boundary. The observed mode splitting and intensity variations highlight the intricate interplay between radial and azimuthal dependencies, which are critical for understanding confined spin wave systems. By controlling the external field orientation and exploiting symmetry-breaking interactions, it is possible to tailor the spin wave spectra of thin rings for specific applications. Potential uses include tunable microwave filters and logic devices that leverage the distinct resonance properties of confined spin-wave modes. The observed sensitivity of spin-wave dynamics to external field orientation offers a mechanism for tuning magnonics\index{magnonics} device performance. Additionally, the well-defined mode structures in thin rings make them ideal candidates for applications requiring precise frequency control.

In conclusion, the investigation of spin-wave dynamics in $30$\,nm-thick permalloy rings has revealed a rich spectrum of behaviors arising from the interplay of radial quantization, symmetry-breaking effects, and external field modulation. These findings advance the understanding of spin-wave phenomena in confined geometries and highlight the potential of thin rings as a platform for magnonic applications.

\textbf{Case of thick (100\,nm) rings}. Ferromagnetic rings have axial symmetry similar to dots, but the topological non-equivalence of the shape of dots and rings leads to specific differences of spin-wave spectra even in the case of thin nanoelements\,\cite{Zho17prb}. The additional effects arising due to varying of the thickness of nanoparticles can be explained if we compare dimensions of spin-wave profiles in films and nanoelements with corresponding materials parameters, i.\,e., the characteristic lengths of exchange and dipolar interactions. The main difference between the standing spin waves in continuous films (where wave vectors $k$ are oriented along the normal to the film plane) and in micrometer-sized patterned elements (where $k$ are oriented along the film plane) is the distance between spin-wave maximums -- in the former case (thickness, $50$-$100$\,nm) it is comparable with material exchange length and in the latter (lateral dimensions, $500$-$2000$\,nm) it is noticeably bigger. However, with further advances in lithography, the difference between the element thickness and lateral dimensions became smaller, and therefore both the exchange and dipolar interactions are starting to play equally important roles. It is expected that the intensity of the spin-wave modes decreases with increase of the mode number due to the diminishing of net magnetization. Recently, it was demonstrated that this simple rule does not apply for the vortex gyrotropic modes in thick circular magnetic dots\,\cite{Zho21prb} due to the complex thickness phase profiles of different spin-wave modes. However, for the spin waves in magnetically saturated samples so far this was always the case.

The study of spin-wave dynamics in thick permalloy rings with a $100$\,nm thickness offers unique insights into the interplay of exchange and dipolar interactions within confined geometries. Unlike thinner rings or continuous films, thick rings exhibit unconventional resonance behaviors that reflect the combined effects of their radial and axial confinement. This section provides an overview of studies  of the standing spin waves in thick rings and discusses their implications for magnonic device design.

The periodic array of permalloy rings was fabricated using a combination of deep ultraviolet lithography, electron beam evaporation, and ultrasonic-assisted lift-off techniques. The process involved the deposition of a $5$\,nm chromium adhesion layer followed by a $100$\,nm permalloy film at a controlled rate of $0.4\,{\AA}$/s under high vacuum conditions (base pressure of $4\times 10^{-8}$\,Torr). The rings were designed with an outer radius of $1500$\,nm and an inner radius of $1100$\,nm, providing a rim width of $400$\,nm. The center-to-center spacing between the rings was $6000$\,nm, ensuring that inter-ring magnetostatic interactions were negligible. An inspection by SEM confirmed structural uniformity across the array. To facilitate comparison, a reference permalloy film with a $100$\,nm thickness was prepared in the same deposition cycle. Its saturation magnetization $M_s$ was found to be $775$\,emu/cm$^3$, as deduced from SQUID measurements.

Broadband perpendicular FMR spectroscopy, as described in the previous section, was employed to investigate the spin-wave dynamics at room temperature. Due to the external magnetic field modulation, the FMR signal detected in this way represents the first derivative of the field sweeping absorption curve at a selected frequency. For each frequency, $H_{op}$ was swept from $18$\,kOe to $0$.

\begin{SCfigure*}
    \centering
    \includegraphics[width=1.48\linewidth]{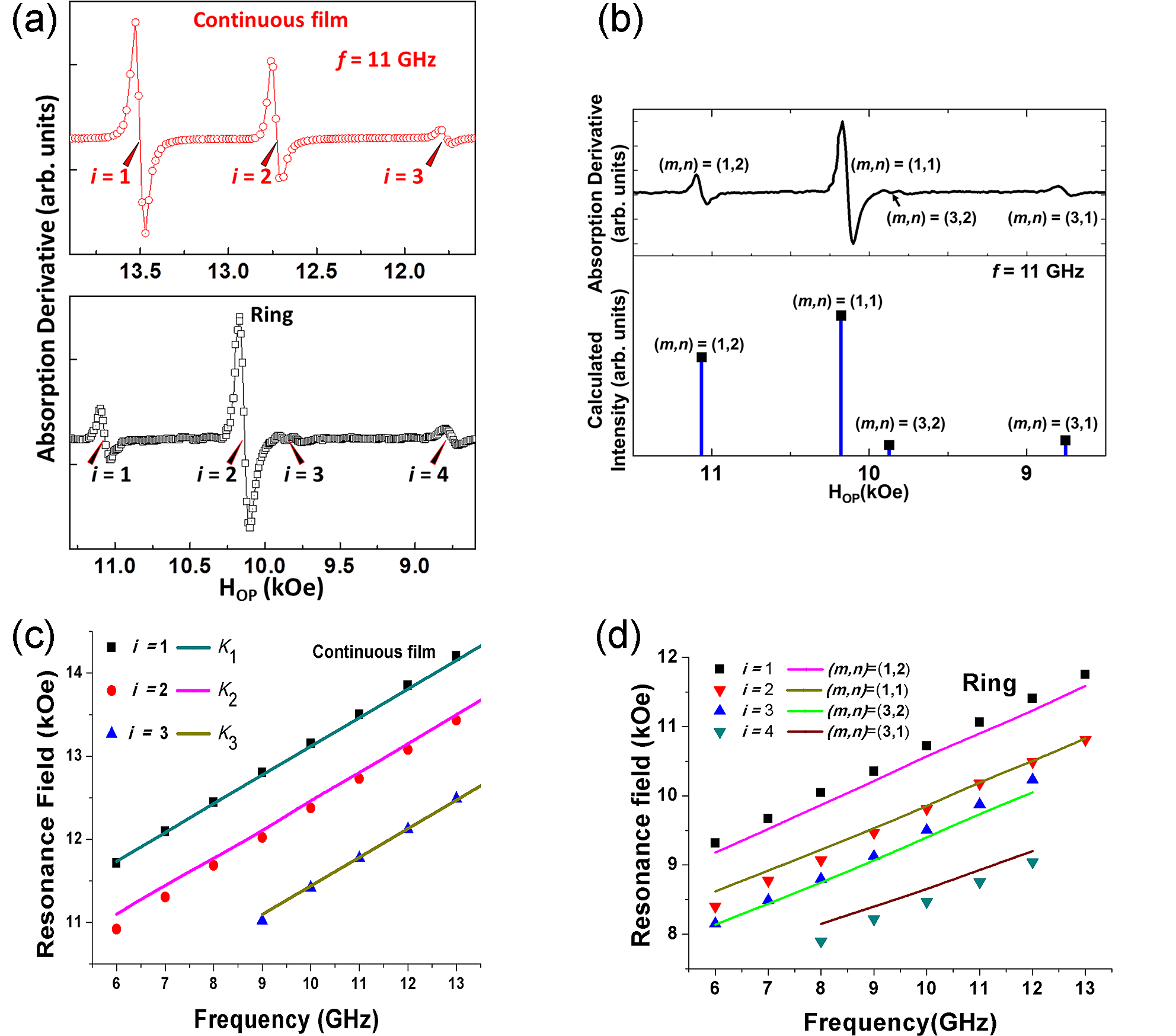}
    \caption{
    (a) Microwave absorption spectra taken at $11$\,GHz for $100$\,nm-thick continuous film (upper panel) and circular rings (lower panel).
    (b) Resonance fields extracted as a function of the excitation frequency for the continuous film: squares, dots and triangles -- experimental results, lines -- calculations by the Kittel formula with the wave vectors $k_1 = 0.47\pi/L$, $k_2 = 1.4\pi/L$, $k_3 = 2.3\pi/L$, where $L$ is the film thickness.
    (c) Resonance fields extracted as a function of the excitation frequency for the circular ring: triangles and squares - experimental results, solid lines -- theoretical calculations by formula (8) from Ref.\,\cite{Zho21prb}. The mode with indices $(m, n) = (1, 2)$ is the highest (magenta) line (experimental black squares), the next (second) lower line corresponds to the mode with indices $(m, n) = (1, 1)$ and has the largest intensity in the FMR experiment (red triangles). The next modes are $(m, n) = (3, 2)$ and $(m, n) = (3, 1)$.
    (d) Calculated intensity of four spin wave modes observed in rings (lower panel) and experimental data for absorption derivative at $11$\,GHz for comparison (upper panel). Adapted with permission from \cite{Zho21prb}.}
    \label{fGleb4}
\end{SCfigure*}

The microwave absorption spectrum for a $100$\,nm-thick permalloy continuous film at $f = 11$\,GHz is shown in the upper panel of Fig.\,\ref{fGleb4}(a). Three distinct resonance modes are clearly observed ($i = 1$: $H = 13.51$\,kOe; $i = 2$: $H = 12.73$\,kOe; $i = 3$: $H = 11.78$\,kOe). The mode intensity gradually decreases with the increasing mode number. This behavior is attributed to the quantization of spin-wave modes along the film's thickness. The spectrum for $100$\,nm thick permalloy rings at the same frequency is shown in upper panel of Fig.\,\ref{fGleb4}(b). In this case, four resonance modes are observed. Different from what was found for the continuous film, for thick rings the first mode ($i = 1$) observed at $H = 11.07$\,kOe shows a smaller absorption amplitude than the second mode ($i = 2$) appearing at $H = 10.14$\,kOe. Besides the two modes and the next intense mode ($i = 4$, $H = 8.76$\,kOe), we also observed an additional mode of small intensity to the right of the second mode ($i = 3$, $H = 9.80$\,kOe).

Shown in Fig.\,\ref{fGleb4}(c) is the experimental dispersion relation ($6$-$13$\,GHz) for the $100$\,nm thick continuous film fitted by the well-known Kittel formula,
\begin{equation}
    f_k=\frac{\gamma}{2\pi}[H_{op}-4\pi M_s+\frac{2A}{M_s}k^2].
\end{equation}
Here $A$ is the exchange stiffness constant and $k$ is the out-of-plane wave vector of the spin-wave excitations. Standard permalloy values were selected for the parameters $A$ and $\gamma$, $A= 1.3\times10^{-6}$\,erg/cm, $\gamma/2\pi = 2.93$\,MHz/Oe, while the quantized values of $k$ were extracted from the experimental results to describe the field gaps between the three observed modes, $k_1 = 0.47\pi/L$, $k_2 = 1.4\pi/L$, $k_3 = 2.3\pi/L$, where $L = 100$\,nm is the film thickness. Such values of $k$ are not standard for spin-wave resonance with ``pinning'' conditions at the surfaces $z = \pm L/2$ (when quantized values of $k$ for detectable modes should be $k_1 = \pi/L$, $k_2 = 3\pi/L$, $k_3 = 5\pi/L$), or for spin wave resonance with ``free'' boundary conditions (when only one homogeneous ``Kittel'' mode with $k_1 = 0$ could be observed in experiment).
Therefore, mixed boundary conditions take place in this case. If $k_2 - k_1$  and $k_3- k_2$ are close to $2\pi$, the equal pinning forces act at the both surfaces of the film, and the boundary conditions for spin wave excitations can be described by only one pinning parameter. However in our case such values are much smaller (actually $k_2 - k_1$ and $k_3- k_2$ are close to $\pi/L$).

This experimental observation gives the evidence of different pinning at the upper and lower film's surfaces, whose strengths can be described by two different pinning parameters, $\sigma_1 = 0.1$ and $\sigma_2 = 10$ (see Ref.\,\cite{Gur96boo} and references therein). Such difference is most probably caused by the fact that permalloy film was deposited on top of Cr underlayer, however no protective layer was deposited on top of permalloy film. As a result, the bottom surface of permalloy film is free from oxide layer (remnant oxygen on the substrate was absorbed by Cr film), therefore pinning is almost absent. On the contrary, the top surface of permalloy film was exposed to the air during chamber venting, which led to its oxidation, and, subsequently, to the noticeable pinning parameter.

It is natural to assume that the boundary conditions for rings are similar to the boundary conditions for the reference film. Hence, the corresponding values of out-of-plane wave vectors $k_i$ are close to the set that was defined before for the $100$\,nm thick continuous film, i.\,e., $k_1 = 0.47\pi/L$ and $k_2 = 1.4\pi/L$. Presented in Fig.\,\ref{fGleb4}(d) is the comparison between the calculation results of the mode frequencies $f_{mn}$ with indices $(m, n) = (1, 1), (1, 2), (3, 1), (3, 2)$ (see Ref.\,\cite{Zho21prb} for more details) and the experimental data. Theoretically calculated frequencies of standing spin waves are in a good agreement with the experimentally observed values for all the four modes. However, the most interesting feature of these calculations is that the mode with indices $(m, n) = (1, 1)$ is excited by a lower resonance field than the mode with indices $(m, n) = (1, 2)$, and, similarly, the resonance field of the mode $(m, n) = (3, 1)$ is lower than of the mode $(m, n) = (3, 2)$. As the intensity of the FMR signal is proportional to the integral by the spatial coordinates from the mode's profile, this means that the mode $i= 1\{(m, n) = (1, 2)\}$ has a smaller intensity than the mode with index $i = 2\{(m, n) = (1, 1)\}$, but a higher resonance field, i.\,e. in comparing with data of the film, rings have the inverse dependence for field vs mode's number $i$.

As the demagnetizing field in infinite films $-4\pi M_s$ does not depend on $K_n$, the field gaps between different modes in the films are defined only by the exchange interactions, i.e. the term $2Ak_n^2/M_s$. Evidently, this dependence provides monotonic decrease of the resonance field with the increase in $k_n$. As opposed to the film, the demagnetizing factors and the dipole-dipole matrix elements in rings are dependent on the mode’s shape. Particularly the mode’s profile along the film thickness and the corresponding wave vector $k_n$ play the most important role. In the case of ring’s geometry dipolar terms ``win this competition'', and modes with larger indices n have smaller frequency at the same excitation field. However, the modes with larger indices n have smaller intensity. This leads to the observed unusual dependence of the FMR signal’s intensity from the number of the mode. Based on the theoretical approach described above, the intensities of four detectable modes of rings using the Eq.\,(11) from Ref.\,\cite{Zho21prb}. The results are presented on the lower panel of Fig.\,\ref{fGleb4}(b) together with the experimental curve of the absorption derivative at $f = 11$\,GHz on the upper panel for comparison. As one may see, there is a fair agreement between the theory and experiment.

The study of thick rings highlights the importance of considering multiple spatial dimensions in analyzing spin wave dynamics. Unlike thin rings, where symmetry-breaking effects dominate, thick rings present a more intricate picture due to the added influence of axial variations. The non-monotonic intensity patterns observed in this study are a direct consequence of the combined effects of radial confinement, axial inhomogeneity, and the competition between exchange and dipolar interactions.

The insights gained from this study have direct implications for the development of magnonic devices. By controlling the thickness and geometry of ferromagnetic rings, it is possible to engineer specific spin wave spectra suited for targeted applications. For example, the non-monotonic intensity patterns observed in thick rings could be exploited to create devices with enhanced sensitivity to specific resonance modes. Additionally, the interplay between exchange and dipolar fields offers a mechanism for dynamically tuning device performance through external field adjustments.

In conclusion, the investigation of spin-wave dynamics in $100$\,nm permalloy rings has revealed a wealth of phenomena arising from the complex interplay of geometric confinement and magnetic interactions. The findings contribute to a deeper understanding of spin-wave behavior in confined structures and highlight the potential of thick rings as a versatile platform for exploring advanced magnonic concepts.

\section{Spin waves in single circular nanomagnets}

\subsection{Spin-wave spectroscopy of individual nanodisks}
\label{cSpectr}

\index{magnetic nanodisks}
Assessing the magnetic parameters of individual nanoelements using local optical probe techniques, such as BLS spectroscopy\,\cite{Seb15fph} and magneto-optical Kerr microscopy\,\cite{Urs16aip}, is challenging due to their spatial resolution limit of approximately $300\,$nm. To investigate magnetization dynamics at GHz frequencies on the scale of a few tens of nanometers, alternative methods are necessary. For example, near-field BLS enables imaging of edge mode dynamics with a resolution of about $50$\,nm\,\cite{Jer10apl}, while X-ray microscopy can achieve dynamic magnetization imaging with a resolution as small as $30$\,nm\,\cite{Slu19nan,Gua20nac}. However, there have been few BLS studies utilizing near-field optics, and X-ray imaging typically requires large-scale facilities. Accordingly, FMR remains the widely used method for quantitatively analyzing saturation magnetization, magnetic anisotropy, dipolar interactions, magnetization relaxation, as well as the structural quality and (in)homogeneity of magnetic materials\,\cite{Far98rpp,Hei91jap,Kal06jap,Kak06prb}.

The sensitivity limitations of traditional FMR\index{ferromagnetic resonance} methods have led to significant advancements aimed at detecting signals from individual magnetic nanoelements. Conventional cavity-based FMR can detect signals from approximately $\simeq10^{10}$ spins, while broadband strip-line FMR can detect signals from around $10^{8}$--$10^{9}$ spins\,\cite{Poo83boo}.
An enhancement of the signal-to noise ratio can usually be achieved for magnetic elements arranged in arrays\,\cite{Ali09prb,Zho15prb}, see Sec.\,\ref{cArrays}. However, an array configuration complicates signal analysis due to inter-element interactions and variations in size and material properties. To address these challenges, more sophisticated FMR-based techniques have been developed, particularly for examining magnetization dynamics in single nanomagnets\,\cite{Nem13prl,Tam14iml,Saf16apl}.
Notable methods include FMR force microscopy (FMRFM) \,\cite{Zha96apl,Lou09prl,Guo13prl,Adu14prl} and microresonator-based FMR (MRFMR) \cite{Mol14sim,Sch17rsi}. Both techniques exhibit enhanced sensitivity of approximately $\simeq 10^6$ spins and have the capability to investigate samples or areas of thin films down to about 100\,nm in size\,\cite{Guo13prl,Sch17rsi,Len19sma}. However, MRFMR requires distinct resonators for each sample, complicating the experimental setup. Likewise, FMRFM faces difficulties in quantitatively analyzing material parameters when the magnetic tip size approaches the size of the sample. Another approach utilizing microwave interferometry has achieved ultrahigh sensitivity, enabling the resolution of uniform-mode FMR signals from single magnetic nanodots as small as $100$\,nm in diameter and $5$\,nm in thickness\,\cite{Tam14iml}. Despite its remarkable sensitivity, this technique is delicate and necessitates the integration of samples directly onto a coplanar waveguide during growth, which adds complexity to the experimental procedure.

An alternative method for achieving large filling factors and high coupling efficiencies involves utilizing microwave strip-lines with narrow and short active components, along with the samples functioning as resonators. The reduced width of the strip-line facilitates the generation of stronger local microwave fields, leading to improved conversion efficiencies from microwave power to field\,\cite{Poz11boo}. Furthermore, the short length of the sensitive section of the strip-line enables spatially-resolved microwave measurements. As a result of the confinement of spin waves, different resonant modes emerge within nanoelements\,\cite{Hil02boo}.
These advancements highlight some of the ongoing efforts to enhance the sensitivity and precision of FMR measurements, paving the way for detailed investigations of magnetic properties at the nanoscale.

This subsection describes a technique for spatially resolved spin-wave spectroscopy to evaluate the magnetic properties of individual circular magnetic elements as small as $100$\,nm in radius. The method utilizes a coplanar waveguide with a $2\times4\,\mu$m$^2$ active area placed over a movable substrate that features well-separated circular magnetic nanoelements. The technique allows for the identification of not only the primary resonance peak but also up to nine higher-order spin-wave excitation modes. By applying an analytical theory, key magnetic parameters, such as saturation magnetization and exchange stiffness, can be determined for each individual magnetic element. Although this approach is adaptable to a variety of magnetic materials, it is particularly beneficial for the quick characterization of individual magnetic nanoelements, especially when their magnetic properties are poorly known.
\begin{SCfigure*}
    \centering
    \includegraphics[width=1.25\linewidth]{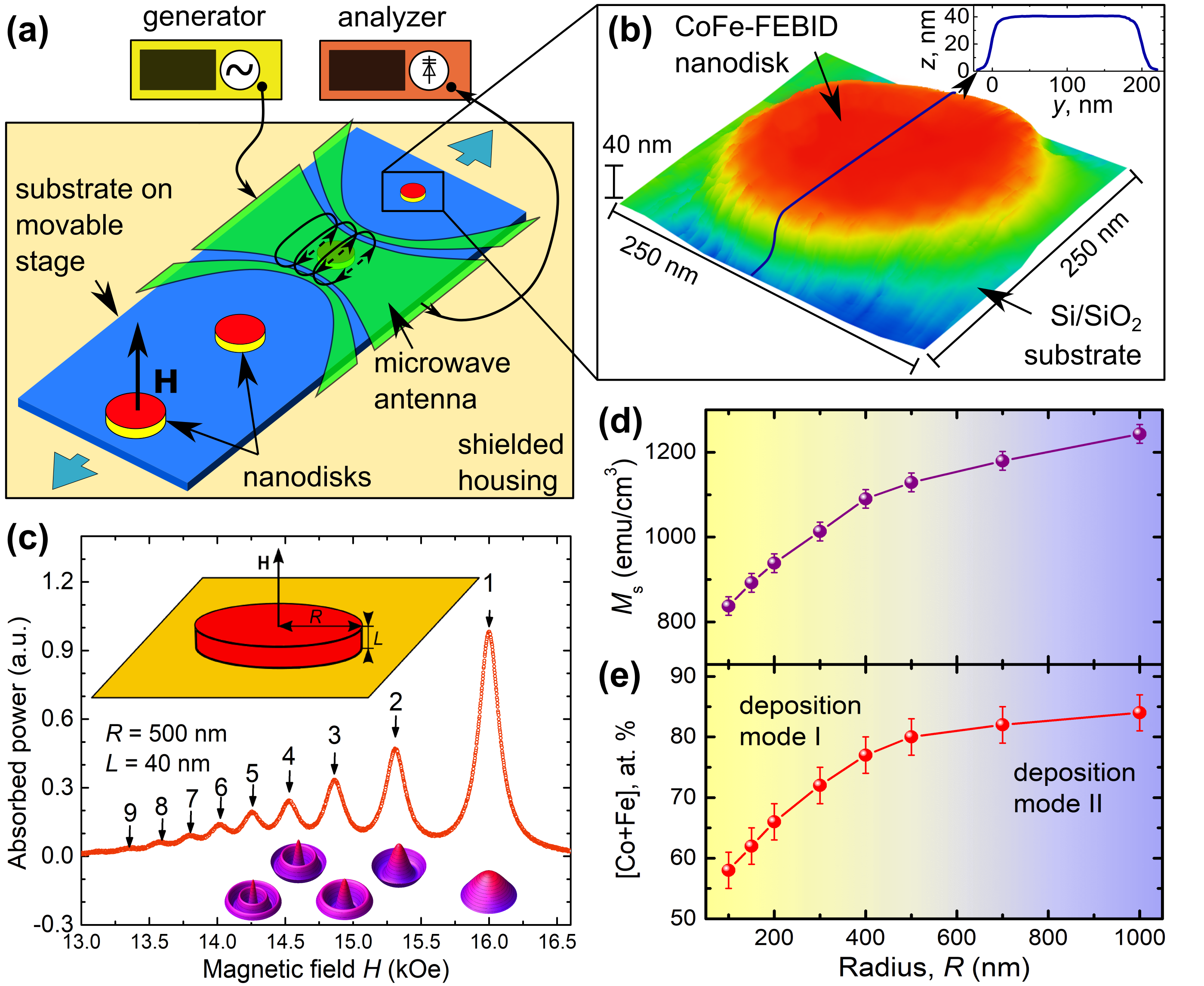}
    \caption{
    (a) A substrate with a series of $40$\,nm-thick CoFe-FEBID nanodisks is positioned directly opposite a gold coplanar waveguide to facilitate spin-wave excitation in a perpendicular-to-disk-plane magnetic field $\mathbf{H}$.
    (b) An AFM image of the smallest nanodisk ($R = 100$\,nm) is shown, with a cross-sectional line scan presented in the inset. (c) The field dependence of microwave power absorption (symbols) is displayed for the disk with a radius of $500$\,nm, revealing at least nine identifiable spin-wave resonance modes, which are labeled near the corresponding peaks. The radial profiles of the first five standing spin-wave modes, described by zeroth-order Bessel functions, are shown. The magnetization $M_{s}$ (d) is derived from the relationship between the resonance field $H_{res}$ and the spin-wave mode number $n$, assuming a gyromagnetic ratio of $\gamma/2\pi = 3.05$\,MHz/Oe. This is compared with the metal content in the nanodisks as determined by EDX spectroscopy (e). The background colors in (d) and (e) signify a transition from the deposition of samples in a depleted-precursor mode (deposition mode I, yellow) to an almost depletion-free precursor mode (deposition mode II, blue) as the disk size increases. Adapted with permission from\,\cite{Dob20nan}.}
    \label{fspectr}
\end{SCfigure*}

The experimental setup is depicted in Fig.\,\ref{fspectr}(a). The samples consist of circular Co-Fe nanodisks, each with a thickness $L$ of $40$\,nm and a radius $R$ ranging from $100$ to $1000$\,nm. The samples were fabricated using focused electron beam induced deposition (FEBID) with HCo$_3$Fe(CO)$_{12}$ as the precursor gas\,\cite{Por15nan}. The composition of [Co+Fe] is approximately 83\,at.\% in the largest disk, tapering down to about 60\,at.\% in the smallest disk. The remaining material consists of carbon and oxygen, remnants from the precursor HCo$_3$Fe(CO)$_{12}$\,\cite{Rag18jpc}. An atomic force microscopy (AFM) image of the nanodisk with $R = 100$\,nm can be seen in Fig.\,\ref{fspectr}(b). The disks were produced in high-resolution deposition mode, utilizing a beam energy of $5$\,keV beam energy, $1.6$\,nA beam current, $20$\,nm pitch, and $1\,\mu$s dwell time, with the beam scanning in a serpentine pattern. Additionally, insulating TiO$_2$-FEBID pads, each measuring $10\times10\,\mu$m$^2$ and 50\,nm in height, were deposited on both sides of the nanodisk row to serve as mechanical spacers for added protection.

The spin-wave excitations in the nanodisks were investigated by positioning the samples on a substrate in close contact with another substrate that had a coplanar waveguide\index{coplanar waveguide} fabricated using e-beam lithography, which was covered with a $5$\,nm-thick insulating TiO$_2$ layer. The substrate containing the antenna was securely attached to a high-frequency sample holder, and precise alignment of the individual nanodisks beneath the sensitive (central) part of the antenna was accomplished using a high-precision piezo stage. Standing spin-wave spectroscopy measurements were conducted at room temperature with a fixed frequency of $f = 9.85$\,GHz, while the magnetic field was oriented perpendicular to the plane of the nanodisks. The measurements were performed in a field-sweep mode, applying a magnetic field modulation $\Delta H$ of 5\,Oe and recording the phase-sensitive derivative of the absorbed microwave power in relation to the magnetic field.

Figure\,\ref{fspectr}(c) shows an exemplary plot of the magnetic field dependence of absorbed microwave power for the disk with $R=500$\,nm. For all nanodisks, the main resonance peak is observed at the largest field (e.g., at $16$\,kOe for the nanodisk with $R=500$\,nm), corresponding to the lowest spin-wave mode number $n = 1$, as will be discussed later. Several distinct resonance peaks appear at lower fields, with their magnitudes decreasing as the mode number increases. As the disk radius decreases, the spin-wave spectra shift to lower fields, and the spacing between neighboring resonances widens.

Standing spin waves are generated in nanoelements due to the quantization of the radial component of the spin-wave vector, which arises from the finite lateral dimensions of the elements\,\cite{Kak04apl}. To analytically determine the field values at which spin-wave resonance occurs, one considers azimuthally symmetric spin waves in a thin cylindrical ferromagnetic disk characterized by a thickness $L$ and radius $R$, which is magnetized in the out-of-plane direction by an external magnetic field $H$ [see the upper inset in Fig.\,\ref{fspectr}(c)]. For circular disks with out-of-plane magnetization, the spin-wave eigenmodes that are excited can be expressed using zeroth-order Bessel functions due to the axial symmetry of the samples \,\cite{Kak04apl}. For further details on the analytical derivations, please refer to Ref.\,\cite{Dob20nan}. The radial profiles for the first nine spin-wave modes are depicted in the lower inset of Fig.\,\ref{fspectr}(c).

The analysis of the experimental relationships $H_{res}(n)$ for the investigated nanodisk series, utilizing the theoretical model\,\cite{Dob20nan}, enables the determination of both the saturation magnetization $M_{s}$ [refer to Fig.\,\ref{fspectr}(d)] and the exchange stiffness $A$ for each individual nanodisk. During the fitting process, $M_{s}$ and $A$ are treated as fitting parameters, with the gyromagnetic ratio fixed at $\gamma/2\pi = 3.05$\,MHz/Oe\,\cite{Tok15prl}. The position of the primary resonance peak is primarily influenced by $M_{s}$ and the sample's demagnetizing factor. While $A$ has a minimal impact on the location of the main resonance peak, it significantly influences the positions of the higher-order peaks. The obtained values for $M_{s}$ indicate that, unlike samples created through electron-beam lithography, the $M_{s}$ of nanostructures produced by FEBID is notably dependent on their lateral dimensions. For instance, the largest nanodisk with $R = 1000$\,nm exhibits $M_{s}$ and $A$ values that are approximately $35\%$ and $10\%$ lower, respectively, compared to the corresponding values for Co$_3$Fe thin films\,\cite{Sch17prb}.

The observed reduction of $M_{s}$ with decreasing disk radius is a result of the writing strategy in the FEBID process and highlights the significance of the characterization detailed here for as-fabricated FEBID samples. Specifically, in the FEBID method, each nanodisk is defined as a circular polygon, where the number of beam dwell points is proportional to the square of the nanodisk radius. The fabrication of the $40$\,nm-thick nanodisks required several thousand passes of the electron beam. Consequently, since the point-to-point distance (pitch) was kept constant for all nanodisks, the electron beam rastered over the smaller disks more quickly than over the larger ones. Due to the finite time required for the precursor gas to replenish in the area around the sample, the smaller disks were created in a depleted-precursor environment. This leads to an increased carbon and oxygen content at the expense of cobalt and iron. The transition from deposition in a depleted-precursor mode (deposition mode I) to a nearly precursor depletion-free mode (deposition mode II) as the disk size increases is depicted by the gradient from yellow to blue in the background of Fig.\,\ref{fspectr}(d).

The concentration of [Co+Fe] in the nanodisks was observed to decline from approximately $83$\,at.\% in the largest disk to around $58$\,at.\% in the smallest disk, while the content of [C+O] increased from about $17$\,at.\% to approximately $42$\,at.\% [see Fig.\,\ref{fspectr}(e)]. Notably, the deduced values of $M_{s}$ in Fig.\,\ref{fspectr}(d) show a strong correlation with the [Co+Fe] content depicted in Fig.\,\ref{fspectr}(e). The correlation suggests that the decrease in $M_{s}$ within the samples can be linked to the reduced presence of the magnetic elements, cobalt and iron, in the volume of the samples.

In summary, this subsection has demonstrated the use of spin-wave spectroscopy on individual circular magnetic elements with radii as small as $100$\,nm and total volumes down to $10^{-3}\,\mu$m$^3$. In this method, the sample functions as a multi-mode resonator, which differs from previous FMR studies that focused on arrays of numerous elements or the examination of single nanoelements within non-uniform magnetic fields using microresonator-based FMR. The circular symmetry of the nanodisks facilitated the accurate determination of the saturation magnetization and the exchange stiffness of the material through an analytical theory. It has been shown how standing spin waves serve as magneto-dynamic probes for nanodisks fabricated via FEBID, highlighting the observed decrease in $M_{s}$ as the disk radius diminishes, attributed to reduced metal content with decrease of the elements' sizes. By adopting an alternative writing strategy, where the electron beam was “parked” for $10$\,ms outside the structure after each pass to allow for precursor replenishment, $M_{s}$ values nearing $1350$\,emu/cm$^3$ have been obtained for a reference set of samples, regardless of nanodisk size, as will be elaborated in the next Sec.\,\ref{cEngin}.

\subsection{Engineered magnetic parameters in nanodisks}
\label{cEngin}

\index{magnetic nanodisks}
The primary challenges in magnonics\index{magnonics} involve the guidance and control of spin waves in one-dimensional (1D) systems, such as magnonic crystals\,\cite{Kra14pcm,Kak14apl,Chu17jpd,Zak20pcm}, two-dimensional (2D) systems like magnonic circuits\,\cite{Mah20jap,Wan20nel}, and the emerging three-dimensional (3D) systems\,\cite{Kra08prb,Yan11apl,Gub19boo}. To steer spin waves, it is necessary to modify an external parameter, which could include the magnetic field\,\cite{Chu09jpd,Liu13apl,Dob19nph}, the temperature\,\cite{Pol12apl,Dzy16apl,Vog18nsr}, or changes to the shape of the conduit\,\cite{Kra14pcm,Gru18prb,Dob19ami} along with adjustments to the magnetization\,\cite{Bau18apl,Dav15prb,Vog18nsr,Whi19prb,Bal14nal,Urb18apl}. Among these methods, varying magnetization offers the advantage of being passive (requiring no current or heat) and can be effectively localized or tailored through gradients as needed. Therefore, there is a strong demand for \emph{in-situ} techniques that allow for magnetization tuning across a wide range. In this context, the impact of ion irradiation on the evolution of the magnetic properties of thin films and nanostructures has been the focus of significant research\,\cite{Har87jap,Ozk02jap,Lan14prb,Rua18aip,Urb18apl,Waw18pra,Ahr20mmm}. The composition and magnetic characteristics of FEBID nanostructures can be adjusted through post-growth irradiation with ions\,\cite{Lar14apl,Dob19ami} and electrons \cite{Beg15nan,Dob15bjn}. FEBID nanostructures are also used for control of the magneto-resistive response in the mixed state of superconductors\,\cite{Dob10sst,Kom14jap,Dob20pra}.

In the previous section, Sec.\,\ref{cSpectr}, we noted that both the saturation magnetization $M_{s}$ and the exchange stiffness $A$ decrease with the reduction of the diameter of individual Co-Fe nanodisks\,\cite{Dob20nan}. This phenomenon is attributed to the fabrication of smaller disks in a depleted-precursor regime, resulting in a decreased metal content. In this subsection, we propose introducing a beam waiting time outside the deposited structures to facilitate the on-demand engineering of magnetization and exchange stiffness in individual direct-write structures. The approach is demonstrated for Co-Fe nanodisks with a thickness of $40$\,nm and a larger fixed radius $R = 500$\,nm.

We present here the results from a series of nanodisks created using varying e-beam waiting times during the FEBID process, as well as another series subjected to different doses of Ga ions. The magnetization $M_{s}$ and exchange stiffness $A$ of these disks are derived from spin-wave resonance (SWR) measurements, utilizing the spatially-resolved technique described in Sec.\,\ref{cSpectr}. Our findings indicate that as the e-beam waiting time increases, the $M_{s}$ of the disks reaches $1430$\,emu/cm$^3$, which is double that of the disks treated with Ga ions. Thus, the integration of these two methods facilitates the creation of geometrically uniform magnonic conduits, enabling significant variation in saturation magnetization.

\begin{SCfigure*}
    \centering
    \includegraphics[width=1.35\linewidth]{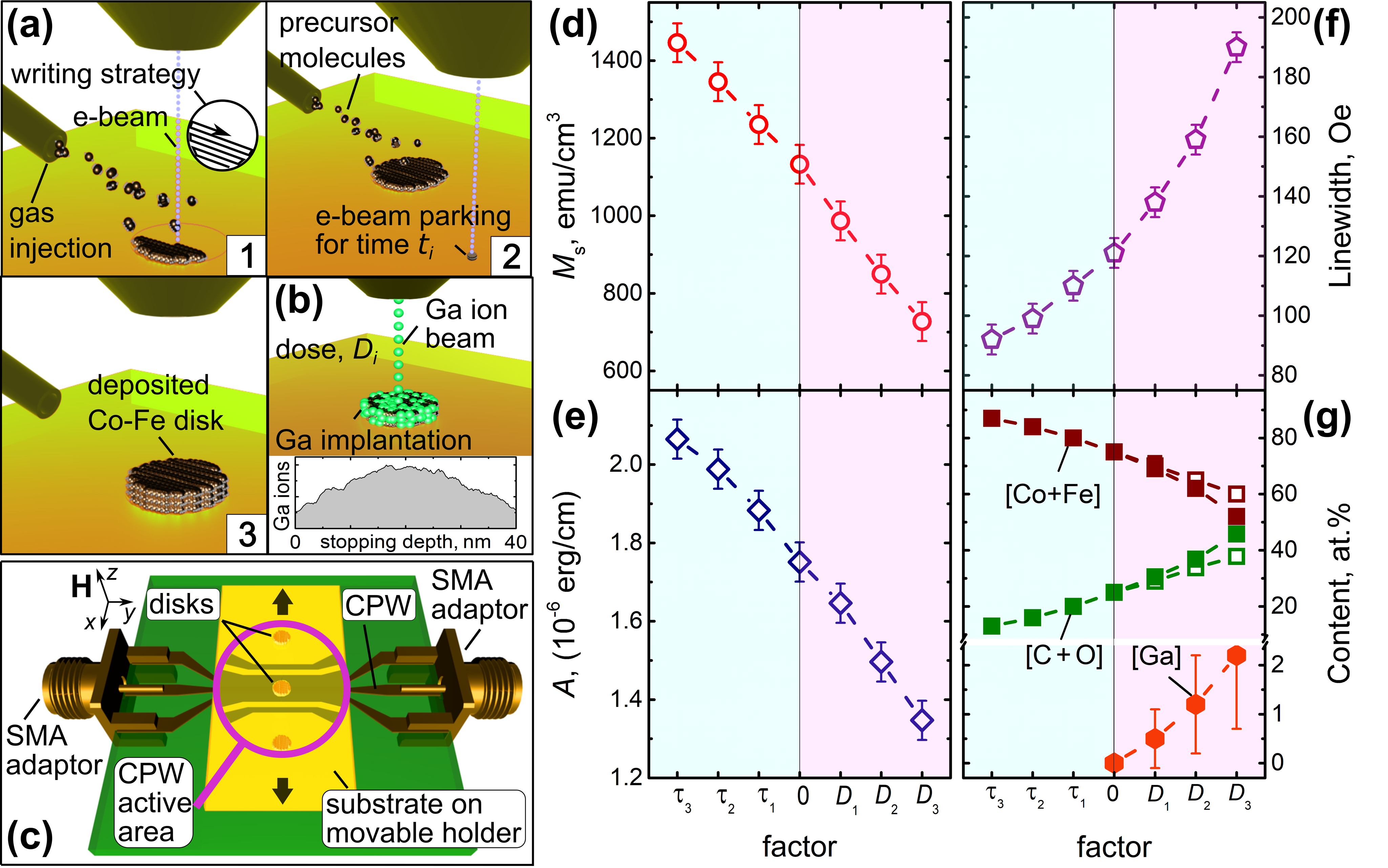}
    \caption{(a) Stages of the FEBID process for the first series of disks: After each pass of the e-beam over the sample surface (1), the beam is parked outside the disk for the time $\tau_i$ (2). This writing process continues until the desired disk thickness is achieved (3).
    (b) In the second measurement series, a Co-Fe disk is subjected to irradiation by $30$\,keV Ga ions at varying doses $D_i$. Inset: Simulated distribution of stopped Ga ions throughout the disk thickness.
    (c) A substrate containing a series of Co-Fe nanodisks is positioned face-to-face with a gold coplanar waveguide for the excitation of spin waves in an out-of-plane bias magnetic field $\mathbf{H}$.
    Evolution of the magnetization $M_{s}$ (d), exchange constant $A$ (e), linewidth (f), and disk composition (g) in relation to increasing electron beam waiting time ($\tau_1$ to $\tau_3$) and Ga ion irradiation dose (doses $D_1$ to $D_3$). Dashed lines serve as guides for the eye. Adapted with permission from\,\cite{Bun21apl}.}
    \label{fengin}
\end{SCfigure*}

The first sample series consists of four disks deposited on a Si/SiO$_2$\,(200\,nm) substrate, each created with varying beam waiting times. After each scanning pass of the electron beam over the disk surface, the beam was ``parked'' for a duration, $\tau$, that ranged from $\tau_0 = 0$ to $\tau_3 = 50$\,ms outside the disk. Figure\,\ref{fengin}(a) illustrates the key steps of the writing process. The thickness variation among the disks written with different $\tau_i$ values was not exceeding $0.5$\,nm. The introduction of the beam ``parking'' time has resulted in a increase of the duration of the writing process by a few minutes.

The substrate was mounted on a translational stage to ensure precise alignment with the $2\,\mu$m-wide and $6\,\mu$m-long active region of an Au coplanar waveguide (CPW), as illustrated in Fig.\,\ref{fengin}(c). The CPW was fabricated using e-beam lithography on a $55$\,nm-thick Au film that was deposited onto a Si/SiO$_2$\,(200\,nm) substrate covered with a $5$\,nm-thick Cr buffer layer. To electrically insulate the disks from the CPW, a $5$\,nm-thick TiO$_2$ layer was applied over the CPW. SWR measurements for both sample series were conducted at a fixed frequency of $9.85$\,GHz, with the magnetic field oriented perpendicular to the plane of the disks, as shown in Fig.\,\ref{fengin}(c).

The second series of samples consists of four different state of a same disk that was initially written with a waiting time of $\tau_0=0$ on the coplanar waveguide (CPW). The disk was subsequently irradiated with $30$\,keV Ga ions up to a cumulative dose $D_3$ of $15$\,pC/$\mu$m$^2$ in steps of $5$\,pC/$\mu$m$^2$, as illustrated in Fig.\,\ref{fengin}(b). SRIM simulations (The Stopping and Range of Ions in Matter\,\cite{Srim}) show that the distribution of $30$\,keV Ga ions within the Co-Fe disks has a gentle dome shape that extends throughout the entire thickness of the disk, with the highest concentration of stopped Ga ions found between depths of $13$\,nm and $28$\,nm, as depicted in the inset of Fig.\,\ref{fengin}(b). As a result of the ion irradiation, the disk thickness reduced to $36.8\pm0.5$\,nm for $D_3 = 15$\,pC/$\mu$m$^2$, which was also accompanied by an increase in surface roughness.

To analytically describe the field values at resonance peaks, azimuthally symmetric spin-wave modes were examined for a thin cylindrical disk that is magnetized in the out-of-plane direction by an applied biasing magnetic field $H$. This method enables the precise determination of both the saturation magnetization $M_{s}$ and the exchange constant $A$\,\cite{Kak04apl}.

The obtained values for $M_{s}$ and $A$ are illustrated in Fig.\,\ref{fengin}(d,e). The resonance linewidth during the field sweep is shown in Fig.\,\ref{fengin}(f). We will now discuss how these values evolve in relation to the disk composition determined through energy-dispersive X-ray (EDX) spectroscopy. The EDX was done at $3$\,kV and $1.6$\,nA, corresponding to a disk thickness of approximately $35$\,nm for the X-ray-emitting layer, as estimated by Monte Carlo simulations (Casino)\,\cite{Dro07sem}. Although the probed layer thickness is expected to be less than the disk thickness in all cases, the open symbols in Fig.\,\ref{fengin}(g) indicate the corrected data that account for a potential oxygen loss from the substrate, which is estimated to be $+3$\,at.\%  after each irradiation step. The EDX results displayed in Fig.\,\ref{fengin}(g) show an increase in the [Co+Fe] content from around $75\%$ at. in the initial sample ($\tau_0=0$) to approximately$87\%$ in the sample written with a beam parking time of $\tau_3=50$\,ms. This increase in metal content corresponds to rises in both $M_{s}$ and $A$, as well as a decrease in linewidth, as shown in Fig.\,\ref{fengin}(f). Conversely, the irradiation with Ga ions leads to a deterioration of the magnetic properties of the nanodisks, resulting in a decline in both $M_{s}$ and $A$, alongside an increase in the linewidth.

The specific values of $\tau$ and $D$ were selected as scale factors in Fig.\,\ref{fengin}(d-g) to illustrate the contrasting nature of the approaches used, as well as to showcase the full range of tuning available for $M_{s}$ and $A$ in Co-Fe nanostructures. The data presented in Fig.\,\ref{fengin} indicate that $M_{s}$ can be adjusted by approximately a factor of two, providing considerable versatility for applications such as the design of graded-index magnonic conduits\,\cite{Dav15prb,Whi19prb,Gru18prb} and magnonic crystals\,\cite{Kra14pcm,Chu17jpd,Zak20pcm}. The findings in Fig.\,\ref{fengin}(f,g) reveal that a reduction in metal content within the disks by roughly $35$\,at.\% corresponds with a two-fold broadening of the linewidth. It is noteworthy that the linewidth of $90$\,Oe at $9.85$\,GHz in the CoFe-rich disk is about twice that observed in sputtered Py films\,\cite{Kal06jap}.

The observed decrease in $M_{s}$ and $A$ in the irradiated disks is attributable to the degradation of ferromagnetic properties due to ion irradiation\,\cite{Har87jap,Ozk02jap,Lan14prb,Rua18aip,Urb18apl,Waw18pra,Ahr20mmm}. It is important to note that ion irradiation can alter the microstructure of the original material, potentially resulting in changes to lattice parameters, grain sizes, and the formation of new phases\,\cite{Ozk02jap}. Although AFM has demonstrated an increase in surface roughness due to irradiation, a thorough microstructural analysis of ion-irradiated Co-Fe will be necessary for future studies.

In summary, we have experimentally revealed an increase in the magnetization $M_{s}$ and exchange stiffness $A$ in disks subjected to longer electron beam waiting times, while a decrease in both $M_{s}$ and $A$ was observed in disks irradiated with Ga ions. This reduction in $M_{s}$ and $A$, along with an increase in linewidth, indicates a degradation of the magnetic properties and greater inhomogeneity in the Ga ion-irradiated disks. Notably, the variation of $M_{s}$ from approximately $720$\,emu/cm$^3$ to about $1430$\,emu/cm$^3$ enables broad engineering capabilities, bridging the gap between the $M_{s}$ values of commonly used magnonic materials\index{magnonic materials} such as Py and CoFeB\,\cite{Chu17jpd}. Correspondingly, the tuning of $M_{s}$ is linked to a change in exchange stiffness, ranging from $1.35\times10^{-6}$\,erg/cm to $2.07\times10^{-6}$\,erg/cm, with field-sweep ferromagnetic resonance linewidths varying between $190$\,Oe and $90$\,Oe. This approach paves the way for developing nanoscale 2D and 3D systems with controllable and spatially varying magnetic properties.

\subsection{Spin-wave eigenmodes in 3D nanovolcanoes}
\label{cVolcano}

\index{magnetic nanovolcanoes}
The exploration of 3D nanomagnets has become an active area of research in magnetism\,\cite{Fer17nac,Fer20mat}, embracing the studies of 3D frustrated systems\,\cite{Kel18nsr,May19cph,Gli20apl,Skj20nrp}, the effects of topology and curvature in intricately shaped nano-architectures\,\cite{Str16jpd,Vol19prl,San20acs,She20cph}, and the dynamics of spin waves within 3D magnonic systems\,\cite{Kra08prb,Yan11apl,Ota16prl,Gub19boo,Sak20apl}. In the realm of magnonics,\index{magnonics} which focuses on the manipulation of data transmitted by spin waves, magnonic conduits have typically been fabricated from 2D structures\,\cite{Kru10jpd,Dem13boo,Liu13apl,Kak14apl,Wan20nel,Liy20jap}. Expanding these spin-wave circuits into the third dimension is essential for reducing the spatial footprint of magnonic logic gates\,\cite{Wan20nel} and permits, for instance, the directional control of spin-wave beams in graded-index magnonics\,\cite{Dav15prb,Toe16nsr,Gra17prb}. In the past, peculiarities of the lithographic process were used, e.g., for the formation of crowns on the tops of nanodisks \cite{Ste11ras}. However, traditional lithographic methods insufficiently meet the needs of 3D magnonics, which has led to a growing interest in additive manufacturing nanotechnologies\,\cite{Fer20mat}.

In this subsection, we discuss the spin-wave eigenmodes of individual direct-write Co-Fe nanovolcanoes analyzed through spin-wave resonance (SWR)\index{spin-wave spectroscopy} spectroscopy\,\cite{Dob20nan,Bun21apl}, alongside micromagnetic simulations. Our findings indicate that the microwave response of the nanovolcanoes is significantly distinct from the combined microwave responses of their 2D components, namely nanorings and nanodisks. We illustrate that the ring surrounding the volcano crater enhances the effective confinement of low-frequency eigenmodes beneath the crater, while higher-frequency eigenmodes are restricted to the ring region. By systematically altering the crater diameter by $\pm20$\,nm, we effectively tune the higher eigenfrequencies by approximately $\pm2$\,GHz, without significantly impacting the lowest eigenfrequency. These nanovolcanoes can thus be regarded as multi-mode resonators and serve as 3D building blocks for the field of nanomagnonics.

\begin{SCfigure*}
    \centering
    \includegraphics[width=1.38\linewidth]{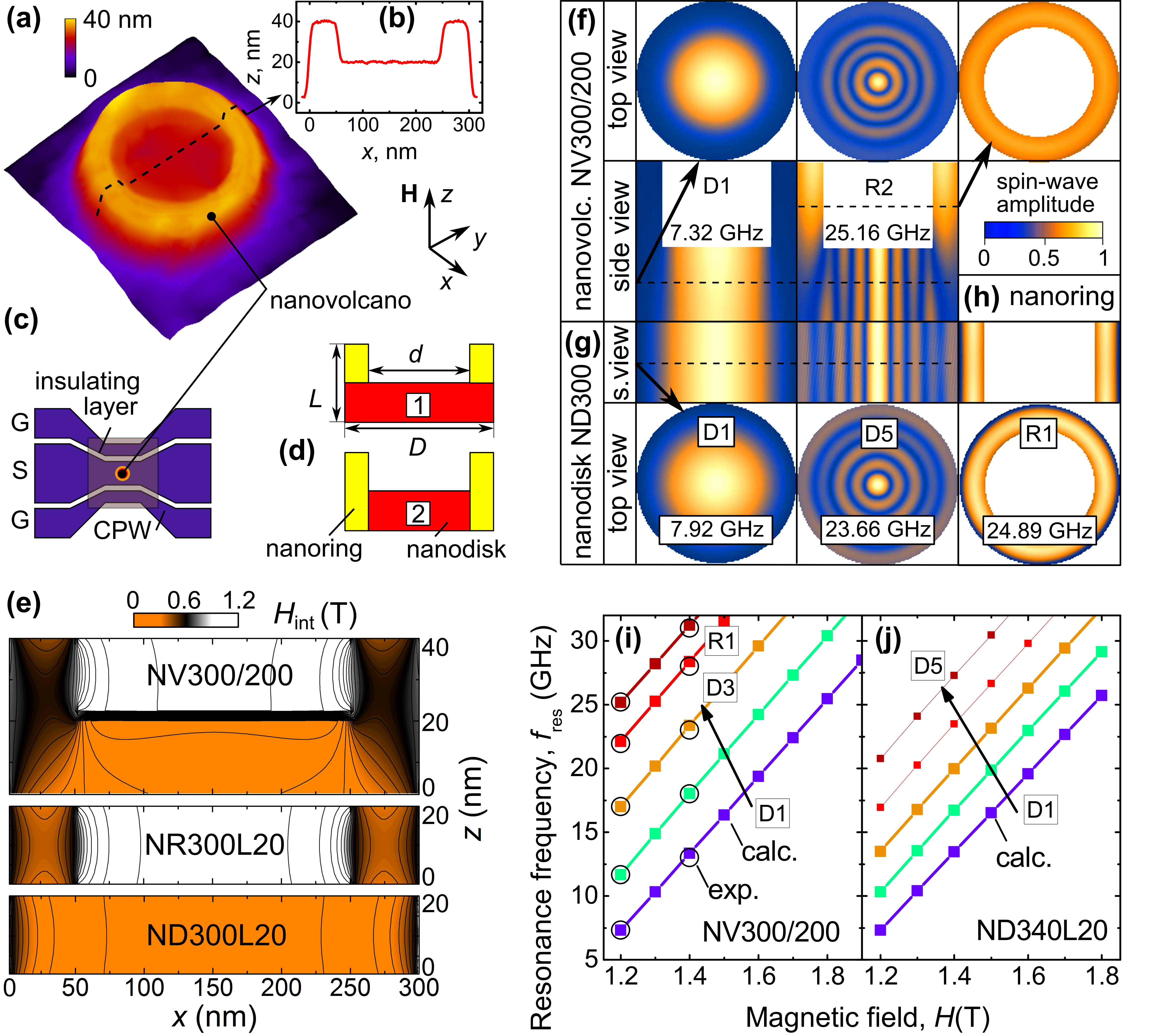}
    \caption{(a) AFM image of the nanovolcano NV300/200, featuring an outer diameter of $D=300$\,nm and a crater diameter of $d=200$\,nm. (b) Cross-sectional line scan. (c) The active part of the coplanar waveguide integrated with the nanovolcano for microwave spectroscopy measurements (not to scale). (d) Potential methods for modeling a nanovolcano as a combination of a disk and a ring. (e) Calculated spatial distribution of the internal field $H_{int}$ for the $40$\,nm-thick nanovolcano NV300/200 and its constituent basic elements (nanodisk ND300L20 and nanoring NR300L20) under an out-of-plane bias magnetic field of $H=1.2$\,T. Calculated spatial distributions of the spin-wave eigenfunctions for (f) the $40$\,nm-thick nanovolcano NV300/200, (g) the $20$\,nm-thick nanodisk ND300, and (h) the $20$\,nm-thick nanoring NR300/200 at $H=1.2$\,T. Resonance frequencies plotted against the bias magnetic field, as obtained from SWR measurements (circles) and numerical calculations (squares) for the nanovolcano NV300/220 (i) and the nanodisk ND340L20 (j). The predominant modes are marked with larger symbols and thicker lines. The slope of the straight lines is given by $\gamma/2\pi=3.05$\,MHz/Oe. Adapted with permission from\,\cite{Dob21apl}.}
    \label{fvolcano}
\end{SCfigure*}

The 40\,nm-thick Co-Fe nanovolcanoes, with outer diameters of $300$\,nm and crater diameters down to $180$\, nm, were created by FEBID on top of a gold CPW, as shown in Fig.\,\ref{fvolcano}(a-c). For the subsequent discussion, we focus on a specific nanovolcano with outer and inner (crater) diameters of $D=300$\,nm and $d=200$\,nm, referred to as NV$D/d$ (with the units ``nm'' omitted). The previously studied simpler structures, nanodisks (ND) and nanorings (NR)\,\cite{Kak04apl,Gie07prb,Sch08prl,Lou09prl,Guo13prl,Lar14apl,Zho17prb} can be considered as the basic components of the nanovolcanoes, see Fig.\,\ref{fvolcano}(d). All nanovolcanoes display a flat morphology and a slightly trapezoidal cross-sectional profile, as revealed by AFM inspection, see Fig.\,\ref{fvolcano}(a,b).

The CPWs were fabricated using e-beam lithography on a Si/SiO$_2$\,(200\,nm) substrate, which had a 55\,nm-thick gold film sputtered onto it atop of a $5$\,nm-thick chromium buffer layer. For electrical insulation from the nanovolcanoes, a $5$\,nm-thick layer of titanium dioxide was deposited using e-beam lithography. The width and length of the narrowed part of the CPW signal line corresponded to $2D$ and $4D$ of the nanovolcano, respectively, see Fig.\,\ref{fvolcano}(c). SWR measurements were conducted over a frequency range of $4$ to $32$\,GHz, with a bias magnetic field from $1.2$ to $2$\,T, oriented perpendicular to the volcano plane. The error in field alignment was less than $0.1^\circ$, small enough to prevent the splitting of spin-wave modes in perpendicularly magnetized nanostructures\,\cite{Bun15nsr}, as discussed in Sec. \ref{cSplit}. A microwave generator provided the high-frequency alternating current excitation, while a signal analyzer was used to detect the transmitted signals. The measurements were carried out with a modulation amplitude of $1$\,mT for the bias magnetic field and at a frequency of $15$\,Hz, employing phase-sensitive recording of the microwave transmission.

The spin-wave spectrum of the nanovolcano features prominent low-frequency spin-wave resonance (SWR) modes (D1) around $7.35$\,GHz, along with distinct higher-frequency SWR modes (R1 and R2). When the crater diameter (d) is increased from $180$\,nm to $220$\,nm, the R1 and R2 peaks exhibit a blue shift of approximately $4$\,GHz\,\cite{Dob21apl}. To identify the SWR modes associated with various components of the nanovolcanoes, micromagnetic simulations were conducted using the MuMax3 solver\,\cite{Van14aip}. The simulation utilized a cell size of $2.5\times$2.5$\times$2.5\,nm$^3$ and a damping parameter set at $\alpha = 0.01$. The simulations were carried out in two phases. Initially, an equilibrium magnetic configuration was established by relaxing a random magnetic arrangement for a specified perpendicular bias field strength. Following this, magnetization precession was stimulated by applying a small, spatially uniform in-plane microwave field pulse. A fast Fourier transform was then utilized to extract the normalized frequency spectra and the spatial distributions of the spin-wave eigenmodes. In these simulations, the saturation magnetization $M_{s}$ and exchange stiffness $A$ values were taken from measurements of disks with identical diameters, fabricated under the same FEBID parameters\,\cite{Dob20nan}, along with the assumed gyromagnetic ratio of $\gamma/2\pi = 3.05$\,MHz/Oe\,\cite{Tok15prl}.

The simulations indicate that the lowest-frequency mode for the nanodisk ND300 occurs at a resonance frequency $f_{res}$ of $7.9$\,GHz [Fig.\,\ref{fvolcano}(g)], which is approximately $600$\,MHz higher than the corresponding experimental value observed for the nanovolcano NV300/200. Conversely, the calculated $f_{res}$ for the 20\,nm-thick nanodisk ND340, which has a diameter of 340\,nm, closely matches the experimentally determined $f_{res}=7.32$\,GHz for the NV300/200 nanovolcano, as shown in Fig.\,\ref{fvolcano}(j). This suggests that the lowest-frequency eigenmode for a model nanovolcano comprised of a $20$\,nm-thick disk (the volcano's base) topped with a $20$\,nm-thick ring can be accurately simulated using a $20$\,nm-thick disk with a larger effective diameter of $D_{eff}=340$\,nm. In other words, extending a 2D nanomagnet into the third dimension allows for the engineering of the lowest eigenfrequency by utilizing a 3D nanovolcano with a footprint that is approximately $30$\% smaller. This phenomenon will be explored in more detail next.

We calculated $f_{res}$ for a $20$\,nm-thick nanoring NR$300/200$ at $1.2$\,T [Fig.\,\ref{fvolcano}(h)]. For the $20$\,nm-thick nanoring NR$300/200$ (geometry 1 in Fig.\,\ref{fvolcano}(d)) $f_{res} = 17.75$\,GHz is very far away from the modes R1 ($22.02$\,GHz) and R2 ($25.16$\,GHz) of the equivalently sized nanovolcano NV$300/200$. This is because of the essentially larger internal field in the volcano's ring area as compared to the $20$\,nm-thick ring (Fig.\,\ref{fvolcano}(e)). By contrast, for a $40$\,nm-thick nanoring NR$300/200$\,nm at $1.2$\,T (geometry 2 in Fig.\,\ref{fvolcano}(d)) $f_{res} = 24.9$\,GHz is very close to the measured R2 eigenfrequency ($25.1$\,GHz) of the equivalently sized nanovolcano NV$300/200$\,nm. Thus, the R2 peak can be attributed to the ring mode of a nanoring that has the same thickness as the nanovolcano and a width corresponding to the ring surrounding the nanovolcano's crater (geometry 2 in Fig.\,\ref{fvolcano}(d)).

The spatial distributions of spin-wave eigenmodes for the nanovolcano NV300/200 are illustrated in Fig.\,\ref{fvolcano}(f). Upon initial examination, the eigenmodes within the nanovolcano are reminiscent of the ``drum modes'' observed in nanodisks with perpendicular magnetic saturation geometry\,\cite{Kak04apl,Dob20nan}. These ``drum modes'' can be roughly represented by zeroth-order Bessel functions\,\cite{Kak04apl} (Eq.\,\eqref{e7}), as shown in Fig.\,\ref{fvolcano}(g) for comparison with the $20$\,nm-thick nanodisk ND300. A detailed analysis of the D1 mode profiles indicates that the spin waves in the nanovolcano are primarily confined beneath the volcano's crater, which acts as a ``concentrator" for spin waves.

We note that the D1 mode frequency for NV300/200 is lower than that for ND300. A qualitative explanation for this observation can be proposed as follows: (i) The internal magnetic field increases at the edge of the NV, causing the D modes to migrate away from this high-field region toward the center of the volcano. Consequently, the stronger confinement of the spin waves results in higher frequencies. This effect becomes more pronounced at higher mode numbers, as the spin-wave frequency is approximately proportional to the square of the wavevector. (ii) In the central region of the NV300/200, the internal field is slightly lower than that in ND300, contributing to a redshift in the D1 frequency. (iii) The influence of the diminished internal field is predominant for the D1 mode, ultimately leading to the observed frequency redshift.

The influence of the ring that overlays the volcano basement disk becomes increasingly significant for the higher-frequency R eigenmodes. This can be attributed to the fact that the diminished effective width, which is proportional to the width $(D-d)/2$ of the ring, becomes more important for shorter wavelengths. The exchange interaction for rings with widths around $50$\,nm plays a key role in determining the R2 frequency (the exchange length \cite{Gus05prb} $L_e = \sqrt{2A/M_s^2}$ in Co-Fe-FEBID is about $5$\,nm). Therefore, the localization of the R2 mode within the ring is not due to reduced dipolar pinning\,\cite{Wan19prl}, but rather arises from the complex interaction between the exchange forces and the internal magnetic fields.

As the magnetic field increases, the entire spectrum shifts towards higher frequencies while maintaining the qualitative structure of the spectra and the relative positions of the peaks. This effect can be attributed to the nearly linear relationship $f_{res}(H)\simeq (\gamma / 2\pi)H$ for axially symmetric nanoelements that are magnetized along their symmetry axis\,\cite{Kak04apl}. In Fig.\,\ref{fvolcano}(i,j), the slopes of all the straight lines representing $f_{res}(H)$ are identical and determined by the gyromagnetic ratio $\gamma / 2\pi$.

In summary, this subsection has introduced nanovolcanoes as nano-architectures for 3D magnetism and magnon spintronics. It has been demonstrated that the spin-wave eigenfrequencies of nanovolcanoes differ significantly from those of the underlying nanodisks and nanorings due to the highly non-uniform internal magnetic field. The experimental observations were clarified using micromagnetic simulations, which reveal that the rings surrounding the volcano craters effectively confine the lower-frequency eigenmodes beneath the crater, while the higher-frequency eigenmodes are localized in the ring area. Consequently, the transition from 2D nanodisks to 3D nanovolcanoes allows for the engineering of their lowest eigenfrequencies while reducing their footprint by approximately 30\%. By varying the crater diameter by $\pm20$\,nm, we demonstrated a tuning capability of around $\pm2$\,GHz for the ring modes, without impacting the lowest spin-wave eigenfrequency. The nanovolcanoes presented can be regarded as multi-mode resonators, offering potential applications in telecom-frequency filters. Furthermore, the engineered spin-wave frequency spectra position them as promising platforms for 3D magnonics and inverse-design magnonic devices.

\section{Conclusion}
Patterned nanostructures are integral not only to industrial applications like magnetic storage but also serve as model systems for fundamental studies of magnetic phenomena. These structures exhibit tunable spin-wave modes, affected by their geometry, material composition, and external magnetic fields. Innovations such as nanovolcanoes have further expanded the potential of these systems, enabling precise control over spin-wave localization and frequency. This combination of experimental advancements and theoretical modeling has broadened the understanding of spin dynamics in confined magnetic systems, laying the groundwork for next-generation technologies in data storage, sensing, and spin-based computing.

The investigation of spin-wave dynamics in magnetic nanostructures, including nanodisks, nanorings, and 3D nanovolcanoes, has provided critical insights into the interplay of geometry, material properties, and external magnetic fields. These studies reveal a wealth of tunable phenomena, from discrete spin-wave modes in nanodisks to localized and frequency-controlled spin waves in nanovolcanoes. The findings underline the potential of these engineered structures to advance magnonic and spintronic applications, offering solutions for efficient information processing and high-density data storage. By bridging experimental observations with theoretical frameworks, this research paves the way for innovative designs in spin-based technologies, contributing to the future of energy-efficient and scalable electronic systems.\\[1mm]

It is our pleasant duty to acknowledge contributions to many of the presented results by our colleagues, A. Adeyeye (Durham University), S. Bunyaev (University of Porto), A. Chumak (University of Vienna), V. Golub (Institute of Magnetism NAN of Ukraine), K. Guslienko (University of the Basque Country), M. Huth (Goethe University Frankfurt am Main), M. Krawczyk (Adam Mickiewicz University Poznan), A. Serga (University Kaiserslautern-Landau), A. Slavin (Oakland University), E. Tartakovskaya (Adam Mickiewicz University Poznan), and X. Zhou (National University of Singapore). O.D. acknowledges financial support from the Austrian Science Fund (FWF) under Grants No. I 4889 (CurviMag) and I 6079  (FluMag) and the Deutsche Forschungsgemeinschaft (DFG, German Research Foundation) under Germany's Excellence Strategy -- EXC-2123 QuantumFrontiers -- 390837967. G.K. acknowledges financial support by FCT -- Portuguese Foundation for Science and Technology through the projects LA/P/0095/2020 (LaPMET), UIDB/04968/2020, UIDP/04968/2020, and 2022.03564.PTDC (DrivenPhonon4Me).


%

\end{document}